\documentclass{emulateapj}
\newcommand{\cmsq}{\mbox{ cm$^{2}$}}
\newcommand{\Mpc}{\mbox{ Mpc}}
\newcommand{\Mpcden}{\mbox{ Mpc}^{-3}}

\newcommand{\erg}{\mbox{ erg}}

\newcommand{\eV}{\mbox{ eV}}
\newcommand{\kel}{\mbox{ K}}

\newcommand{\yr}{\mbox{ yr}}

\newcommand{\secinv}{\mbox{ s$^{-1}$}}

\newcommand{\Msun}{\mbox{ M$_\odot$}}

\newcommand{\Lsun}{\mbox{ L$_\odot$}}

\newcommand{\sfr}{\mbox{ M$_\odot$ yr$^{-1}$}}

\newcommand{\hunits}{\mbox{ km s$^{-1}$ Mpc$^{-1}$}}

\newcommand{\kms}{\mbox{ km s$^{-1}$}}

\newcommand{\recunits}{\mbox{ cm$^{3}$ s$^{-1}$}}

\newcommand{\bxion}{\bar{x}_i}

\newcommand{\hone}{HI }

\newcommand{\mmin}{m_{\rm min}}
\newcommand{\fcoll}{f_{\rm coll}}

\newcommand{\lya}{Ly$\alpha$ }

\newcommand{\deriv}{{\rm d}}
\newcommand{\bq}{\begin{equation}}
\newcommand{\eq}{\end{equation}}
\newcommand{\bqa}{\begin{eqnarray}}
\newcommand{\eqa}{\end{eqnarray}}
\def\VEV#1{\left\langle #1\right\rangle} 

\begin{document}

\title{The History and Morphology of Helium Reionization}

\author{Steven R.  Furlanetto\altaffilmark{1} \& S.~Peng Oh\altaffilmark{2}}

\altaffiltext{1} {Department of Physics and Astronomy, University of California at Los Angeles; Los Angeles, CA 90095; sfurlane@astro.ucla.edu}

\altaffiltext{2}{Department of Physics, University of California, Santa Barbara, CA 93106;  peng@physics.ucsb.edu}

\begin{abstract}
A variety of observations now indicate that intergalactic helium was fully ionized by $z \sim 3$.  The most recent measurements of the high-redshift quasar luminosity function imply that these sources had produced at least $\sim 2.5$ ionizing photons per helium atom by that time, consistent with a picture in which the known quasar population drives \ion{He}{2} reionization.  Here we describe the distribution of ionized and neutral helium gas during this era.  Because the sources were rare and bright (with the photon budget dominated by quasars with luminosities $L \ga L_\star$), random fluctuations in the quasar population determined the morphology of ionized gas when the global ionized fraction $\bxion$ was small, with the typical radius $R_c$ of a \ion{He}{3} bubble $\sim 15$--$20$ comoving Mpc.  Only when $\bxion \ga 0.5$ did the large-scale clustering of the quasars drive the characteristic size of ionized regions above this value.  Still later, when $\bxion \ga 0.75$, most ionizing photons were consumed by dense, recombining systems before they reached the edge of their source's ionized surroundings, halting the bubble growth when $R_c \sim 35$--$40 \Mpc$.  These phases are qualitatively similar to those in hydrogen reionization, but the rarity of the sources makes the early stochastic phase much more important.  Moreover, the well-known characteristics of the $z=3$ intergalactic medium allow a much more robust description of the late phase in which recombinations dominate.
\end{abstract}
  
\keywords{cosmology: theory -- intergalactic medium -- quasars}

\section{Introduction} \label{intro}

For most of the Universe's history, the intergalactic medium (IGM) evolved rather slowly and smoothly.  But there were two major exceptions:  the reionization of hydrogen (at $z \ga 6$) and of helium (at $z \sim 3$).  Recently, hydrogen reionization has received a great deal of attention in both the observational and theoretical communities (see, e.g., reviews by \citealt{barkana01, ciardi05-rev, fan06-review, furl06-review}).  Helium reionization has received less attention, especially from theorists, but has actually been more accessible observationally and has important ramifications for quasar populations, galaxy formation, and the structure of the IGM itself.

Because \ion{He}{2} has an ionization potential of 54.4 eV, the relatively soft photons produced by (known populations of) hot stars do not ionize it to any large degree (they can, on the other hand, singly ionize helium along with hydrogen).  As a result, helium remained singly ionized until the quasar population built up in sufficient numbers:  quasars have hard spectra, with luminosity densities $L_\nu \propto \nu^{-1.6}$ \citep{telfer02} and can produce enough photons to complete the reionization process by $z \sim 3$ \citep{sokasian02, wyithe03, gleser05}.  

Observations do indeed point to helium reionization at $z \sim 3$.  The strongest evidence comes from far-ultraviolet spectra of the \ion{He}{2} \lya forest along the lines of sight to several bright quasars at $z \sim 3$:  the observed wavelength of such an absorption system is $\lambda = 304(1+z)$ \AA.  First suggested as a probe of the low-density IGM \citep{miralda93, giroux95, miralda96, croft97}, the \ion{He}{2} forest has also proven to be a powerful probe of reionization and of the background radiation field \citep{fardal98}.  For our purposes, the most important point is that the apparent \ion{He}{2} optical depth decreases rapidly at $z \approx 2.9$, with a spread of $\Delta z \approx 0.1$ along different lines of sight \citep{jakobsen94, davidsen96, anderson99, heap00, smette02, zheng04, shull04, reimers04, reimers05, fechner06}.  This apparent transition is analogous to the rapidly-increasing \ion{H}{1} optical depth observed at $z \sim 6$ and usually attributed to reionization \citep{fan01, fan06}, though see \citet{becker07} for a different interpretation.  Although appropriate lines of sight are rare (they require a bright quasar with no intervening Lyman-limit absorbers that attenuate the far-UV flux), recent large-scale surveys such as SDSS have opened up many more targets (e.g., \citealt{zheng04-sdss}).

We may thus be able to observe the helium reionization era in detail.  Particularly interesting are the large opacity fluctuations observed along several lines of sight \citep{anderson99, heap00, smette02, reimers05}.  At least one of these coincides with a nearby quasar \citep{dobrzycki91, jakobsen03}.  This illustrates a crucial advantage of studying helium reionization over hydrogen:  we understand the $z \sim 3$ Universe much better than the $z \sim 6$ Universe, in terms of the ionizing sources, the IGM, and other galaxies.  We can thus provide much sharper tests of reionization models.  

There are several other indirect lines of evidence for helium reionization at $z \sim 3$, although in each case they are controversial.  Helium reionization should at least double the IGM temperature.  \citet{schaye00} detected a sudden increase in the IGM temperature at $z \sim 3.3$ by examining the Doppler widths of \ion{H}{1} \lya forest lines (see also \citealt{schaye99, theuns02-reion}); at about the same time, the equation of state of the IGM also appears to become nearly isothermal, another indication of recent reionization \citep{schaye00, ricotti00}.  However, temperature measurements via the \lya forest flux power spectrum show no evidence for any sudden change, although the errors are rather large \citep{zald01, viel04, mcdonald06}.  Recent models of the helium reionization process predict sudden jumps in the temperature as measured by the \citet{schaye00} method, although with somewhat smaller magnitudes, along with smoother evolution in the mean temperature, which may help to resolve the controversy \citep{gleser05, furl07-igmtemp}.

Such a temperature jump should also decrease the recombination rate of hydrogen, thereby decreasing the \ion{H}{1} opacity \citep{theuns02-sdss}.  \citet{bernardi03} detected such a jump at $z \sim 3.2$ in a large sample of SDSS spectra; however, \citet{mcdonald06} found no such feature with similar data and precision.  Most recently, \citet{fauch07} studied a sample of high-resolution \lya forest spectra and found a feature nearly identical to that in \citet{bernardi03}.

Finally, one would also expect the (average) metagalactic ionizing background to harden as helium is reionized, because the IGM would become transparent to high-energy photons.  This should manifest itself in the \ion{He}{2}/\ion{H}{1} ratio (as possibly observed along one line of sight; \citealt{heap00}) and also in optical spectra.  In particular, the ionization potentials of \ion{Si}{4} and \ion{C}{4} straddle that of \ion{He}{2}, so their ratio should evolve during helium reionization.  \citet{songaila98, songaila05} found a break in their ratio at $z \sim 3$; modeling of the ionizing background from optically thin and optically thick metal line systems also shows a significant hardening at $z \sim 3$ \citep{vladilo03, agafonova05, agafonova07}.  However, other data of comparable quality show no evidence for rapid evolution \citep{kim02, aguirre04}.  This approach is made more difficult by the large fluctuations in the \ion{He}{2}/\ion{H}{1} ratio even after helium reionization is complete \citep{shull04}.

Despite this wealth of (often controversial) data, helium reionization has received relatively little theoretical attention.  There have been a few attempts to estimate the evolution of the globally-averaged \ion{He}{3} fraction with redshift, given models for the ionizing sources as input (e.g., \citealt{wyithe03}).  There has also been one numerical simulation of helium reionization \citep{sokasian02}, which confirmed that quasars provided enough photons to complete the process at $z \sim 3$--$4$ and that the evolving opacity of the \ion{He}{2} \lya forest was consistent with the tail end of reionization.  Most recently, \citet{gleser05} used a semi-analytic Monte Carlo model to examine both the optical depth and temperature evolution in the context of helium reionization.  However, there have been no efforts to understand the fundamental question of how the ionized and neutral gas are organized in the IGM (the ``morphology" of reionization), which provides the overall paradigm in which observations must be interpreted.

On the other hand, in the past several years hydrogen reionization has received an enormous amount of theoretical attention, including both analytic models and simulations (see \citealt{furl06-review} for a recent summary).  In particular, we now appreciate the importance of source clustering for reionization, which makes the process inhomogeneous on extremely large scales ($\ga 10 \Mpc$) and substantially affects the interpretation of many observables (\citealt{barkana04, furl04-bub}, hereafter FZH04).  The purpose of this paper is to apply the simplest of these models to the helium reionization epoch as a first step toward understanding the large-scale inhomogeneity of the process.  In particular, we will describe three distinct stages in the evolution of the morphology, and we will emphasize the similarities and differences in the two eras.  We base our models on FZH04 and \citet[hereafter FO05]{furl05-rec}.

This paper is organized as follows.  We begin in \S \ref{history} by considering simple reionization histories.  We then describe stochastic and clustering-driven models for the bubble sizes in \S \ref{stochastic} and \S \ref{bub}, respectively.  We next consider the role of inhomogeneous recombinations in \S \ref{recomb}.  Finally, we conclude in \S \ref{disc}.

In our numerical calculations, we assume a cosmology with $\Omega_m=0.26$, $\Omega_\Lambda=0.74$, $\Omega_b=0.044$, $H=100 h \hunits$ (with $h=0.74$), $n=0.95$, and $\sigma_8=0.8$, consistent with the most recent measurements \citep{spergel06}.  Unless otherwise specified, we use comoving units for all distances.

\section{The Reionization History} \label{history}

\subsection{An Empirical Model} \label{empirical}

\begin{figure*}
\plottwo{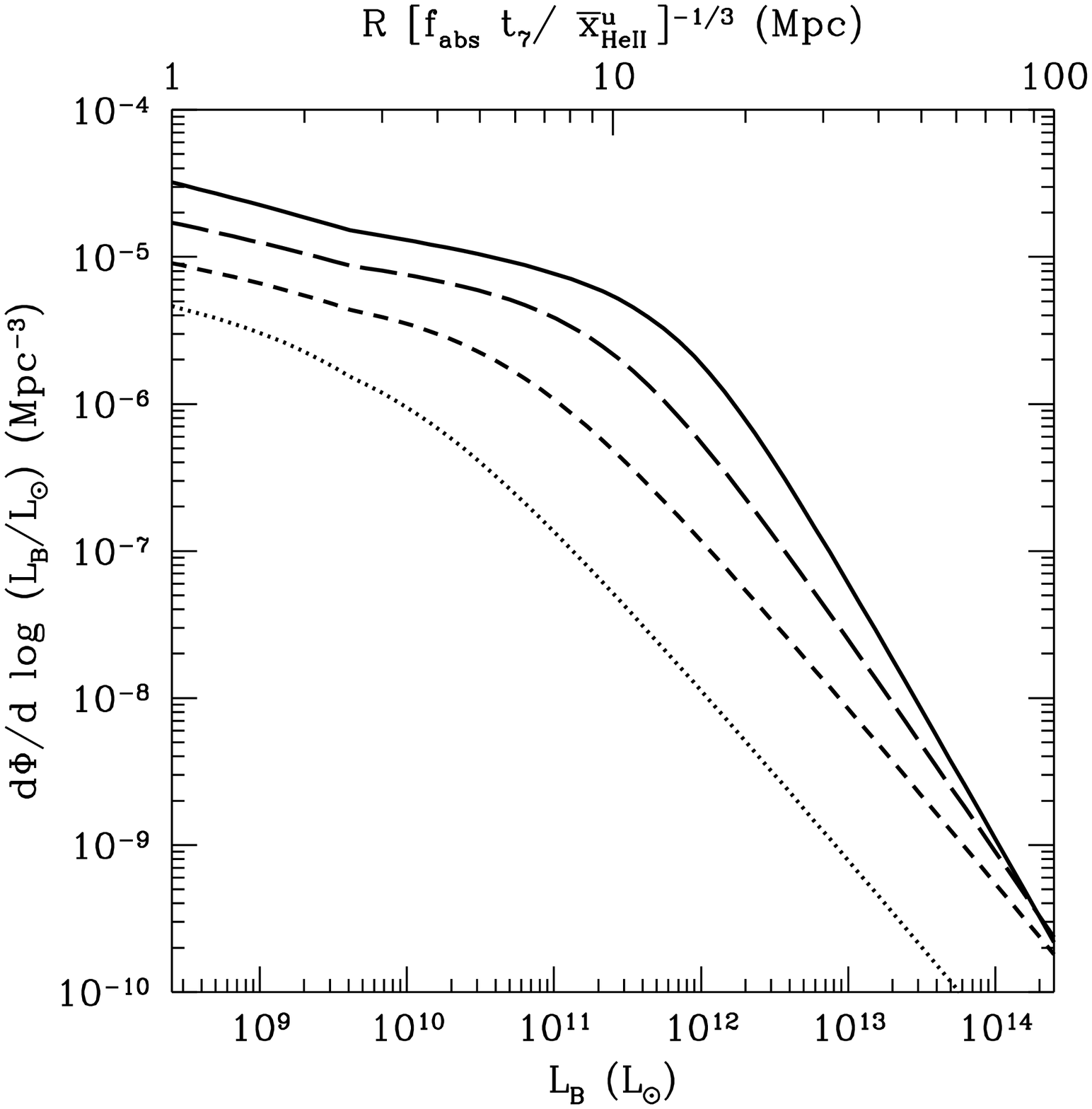}{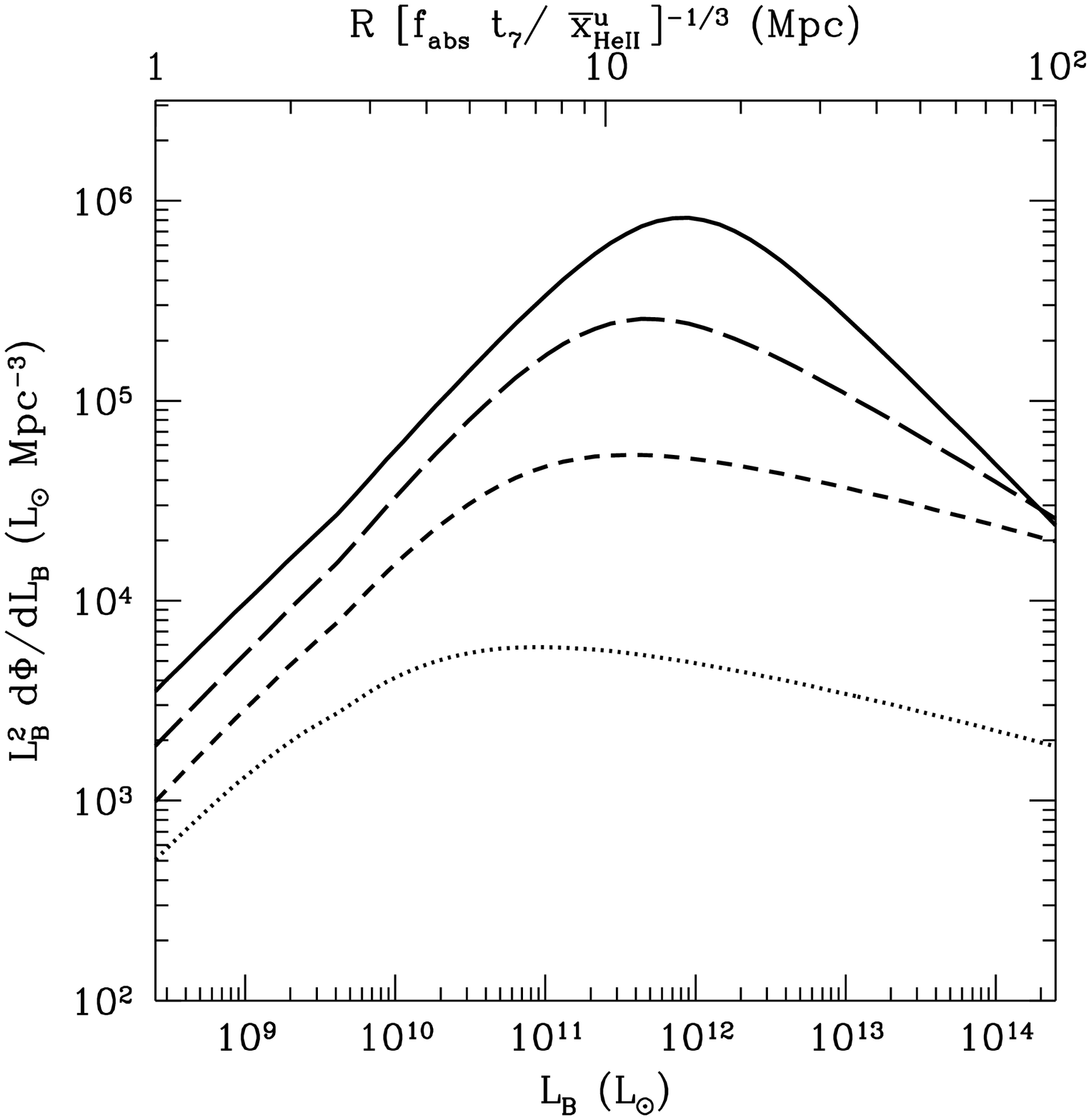}
\caption{\emph{Left panel:}  Quasar luminosity function from \citet{hopkins07}.  \emph{Right panel:}  Same, but weighted by $L_B$.  In each panel, the solid, long-dashed, short-dashed, and dotted curves assume $z=3,\,4,\,5$, and $6$, respectively.  Along the upper axis, we show the bubble size (in comoving units) corresponding to each luminosity, parameterized in terms of the fraction of ionizing photons absorbed inside the bubble ($f_{\rm abs}$), the quasar lifetime in units of $10^7 \yr$ ($t_7$), and $\bar{x}_{\rm He II}^u$.}
\label{fig:qlf}
\end{figure*}

We first consider how the globally-averaged \ion{He}{3} fraction, $\bxion$, evolves with redshift.  To begin, we will assume that quasars drive the reionization process.  In that case, the key empirical input to our model is the quasar luminosity function.  
We use the recent estimate of the bolometric luminosity function over a broad range of redshifts from \citet{hopkins07}, who convert to a $B$-band luminosity function using the observed column density distribution of quasars selected in the X-ray (thus accounting for obscured sources).  The distribution of $B$-band luminosities $L_B$ (which we define to be $\nu L_\nu$ evaluated at 4400 \AA, the center of the $B$-band\footnote{Note that we use a different luminosity convention than \citet{sokasian03}.}) is shown in the left panel of Figure~\ref{fig:qlf} for $z=3,\,4,\,5,$ and $6$, from top to bottom.  Here $\deriv \Phi/\deriv \log L_B$ is the number density of quasars per logarithmic interval in luminosity.

The \citet{hopkins07} luminosity functions are consistent with earlier estimates, but they do have significantly flatter faint-end slopes (e.g. $\deriv \Phi/\deriv \log L_B \propto L^{0.4}$ for $L_B \ll 10^{12} \Lsun$ at $z=3$).  In the sample they used, the assumption of pure luminosity evolution no longer fit the data at high redshifts, which at the time showed a flattening at the faint end \citep{hunt04, cristiani04}.  More recent data show a somewhat steeper faint end slope ($\propto L^{0.7}$) at $z \sim 3$ \citep{fontanot07, bongiorno07, siana08}, more consistent with lower-redshift  estimates.  For simplicity we will use the \citet{hopkins07} function throughout, but we will also note where a steeper faint end slope affects our results.  (It is not trivial to simply change the faint end slope in the fit, because it is covariant with other parameters like $L_\star$.)

The right-hand panel of Figure~\ref{fig:qlf} shows $\deriv \Phi/\deriv L_B$ weighted by luminosity, which is proportional to the number of photons produced per logarithmic unit of quasar luminosity.  This panel explicitly pinpoints the quasars that contribute to the ionizing photon budget during reionization.  Clearly, at $z \sim 3$, quasars with $L_B \sim L_\star \approx 10^{12} \Lsun$ dominate.  At higher redshifts, the luminosity function flattens at the bright end; hence, a much broader range of quasar sources contribute significantly.  (However, note that the faint end converges even at $z=6$, so quasars fainter than our assumed minimum quasar luminosity of $10^6 \Lsun$ contribute negligibly.)

We must next convert from these $B$-band luminosity functions to the rate at which the quasars produce helium-ionizing photons with $E>54.4 \eV$:  this requires a template for the spectral energy distribution of quasars.  We use
\bq
L_\nu \propto \left\{ 
\begin{tabular}{ll}
$\nu^{-0.3}$ & \qquad 2500 \mbox{ \AA} $< \lambda <$ 4400 \mbox{ \AA} \\
$\nu^{-0.8}$ & \qquad 1050 \mbox{ \AA} $< \lambda <$ 2500 \mbox{ \AA} \\
$\nu^{-\alpha}$ & \qquad $\lambda <$ 1050 \mbox{ \AA}.
\end{tabular}
\right.
\label{eq:qsotemplate}
\eq
At $\lambda > 1050$ \AA, this template agrees with that of \citet{madau99-qso}; other templates (e.g., \citealt{schirber03}) disagree in detail but do not affect our conclusions, given the uncertainties.  Most important for us is the far-ultraviolet spectral index $\alpha$.  At low redshifts, \citet{telfer02} find a wide variety of quasar spectral indices in the extreme ultraviolet, with a mean value of $\VEV{\alpha} \approx 1.6$ and a standard deviation $\sigma_\alpha \approx 0.8$.  This is slightly harder than the estimate of \citet{zheng98}, who found $\VEV{\alpha} \approx 1.8$.  We will use $\alpha = 1.6$ for most of our calculations, but note that these uncertainties do affect our estimates of the absolute rate at which ionizing photons are produced.

Given the template, the rate at which a quasar produces ionizing photons is
\bqa
\dot{N}_i & = & 0.0948 {L_B \over h} \left( {228 \, \mbox{\AA} \over 1050 \, \mbox{\AA}} \right) ^{\alpha} \int_{\nu_{\rm He}}^\infty {\deriv \nu \over \nu} \left( {\nu \over \nu_{\rm He}} \right)^{-\alpha} \\
& = & 2.0 \times 10^{55} \secinv \left( {L_B \over 10^{12} \Lsun} \right),
\label{eq:nphdot}
\eqa
where in the second line we have assumed $\alpha=1.6$.  (For $\alpha=1.8$, the prefactor becomes $1.4 \times 10^{55} \secinv$.)

To develop intuition, it is convenient to assume a quasar lifetime $t_{\rm QSO} = 10^7 t_7 \yr$.  In that case, a quasar can ionize a region of radius
\bq
R_i \approx 14 \left[ {f_{\rm abs} t_7 \over \bar{x}^u_{\rm HeII}} \,  { L_B \over 10^{12} \Lsun} \right]^{1/3} \Mpc,
\label{eq:Vi}
\eq
where $f_{\rm abs}$ is the fraction of ionizing photons absorbed within the ionized bubble (as opposed to escaping into the surrounding IGM; see below) and the quasar lifetime is $t_{\rm QSO}=10^7 t_7 \yr$.  We also assume that the bubble expands into a uniformly ionized background with ionized fraction $\bxion^u = 1 - \bar{x}_{\rm HeII}^u$.  Along the upper axes of Figure~\ref{fig:qlf}, we show the bubble radius corresponding to each luminosity.  Clearly, we expect the typical bubbles around isolated quasars to be $\sim 10$--20 comoving Mpc across, although clustering will substantially increase the true sizes.

We emphasize that $R_i$ is considerably smaller than the nominal ``proximity zone" around each quasar (i.e., the region in which its ionization rate exceeds that of any other quasar), even if the IGM is fully ionized.  For example, the mean distance between $L_\star$ quasars at $z=3$ is $\ell \sim n_q^{-1/3} \sim 110 \Mpc \gg R_i$.  This is a reflection of the short lifetimes of quasars:  $(R_i/\ell)^3 \sim t_7 H(z=3)$, so that over the age of the universe the entire quasar population can reionize the IGM.  Most \ion{He}{3} regions are not ``active" but are instead fossils, which have been ionized by a now extinct quasar (see \S \ref{fossils}). 

The ionized fraction $\bxion$ will evolve following
\bq
{\deriv \bxion \over \deriv t} = \int \deriv L_B {\dot{N}_i \over \bar{n}_{\rm He}} {\deriv \Phi \over \deriv L_B} - \bar{C} A_u \bxion, 
\label{eq:xevol}
\eq
where $\bar{n}_{\rm He}$ is the mean helium density, $\bar{C} \equiv \VEV{n_e^2}/\VEV{n_e}^2$ is the average clumping factor in the ionized IGM, and $A_u$ is the recombination rate per helium atom in gas at the mean density,
\bq
A_u = \alpha_A(T) \bar{n}_e.
\label{eq:au}
\eq
We use the recombination coefficients from \citet{storey95}, $\alpha_A = 2.18 \times 10^{-12} \recunits$ and $\alpha_B=1.53 \times 10^{-12} \recunits$ at $T=20,000 \kel$.  Unfortunately, the choice between case-A and case-B (which ignores recombinations to the ground state) is not an obvious one \citep{miralda03}.  Many studies use case-B, because any recombinations that regenerate an ionizing photon do not produce a net change in the ionized fraction.  However, in a clumpy universe most recombinations actually occur near dense, mostly neutral systems (Lyman-limit systems for hydrogen-ionizing photons).  Ionizing photons produced in these systems will be absorbed before they can escape into the low-density, ionized IGM for which we wish to compute the recombination rate.  We therefore generally use the case-A rate below.  Throughout this work, we will assume an IGM temperature $T = 20,000 \kel$, appropriate for photoionization with a relatively hard ionizing source.  

For gas at the mean density, the ratio of the helium recombination time to the Hubble time is
\bq
{t_{\rm rec} \over H^{-1}(z)} \approx 0.4 \left( {4 \over 1+z} \right)^{3/2} \left[ {\alpha_A(2 \times 10^4 \kel) \over \alpha} \right],
\label{eq:rectime}
\eq
where we have assumed matter domination.  Because recombinations are so fast, the sink term in equation~(\ref{eq:xevol}) -- and hence the clumping factor -- is quite important.  We will argue below that $\bar{C}$ will increase throughout reionization (FO05), but for the purposes of this simple calculation we will consider a range of constant clumping factors.

\begin{figure}
\plotone{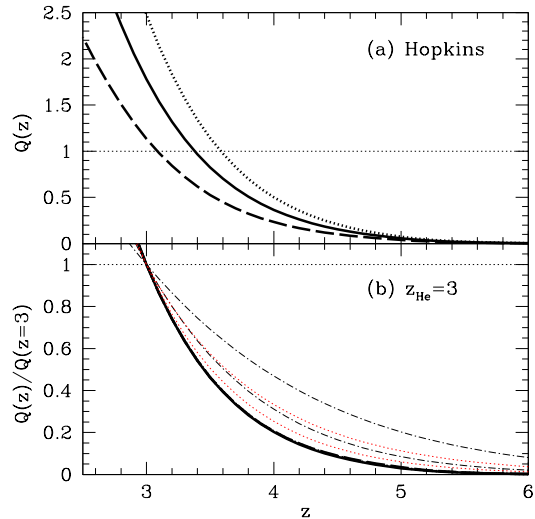}
\caption{Helium reionization histories.  \emph{(a)}:  Using the \citet{hopkins07} luminosity function, assuming $\bar{C}=0,\,1$, and $3$ (dotted, solid, and dashed curves, respectively).  \emph{(b)}:  Histories normalized so that $z_{\rm He}=3$.  The thick curves (which all overlap) are identical to those in \emph{(a)}.  The upper and lower dot-dashed curves follow $\fcoll$ for $\mmin = m_i$ and $10 m_i$, respectively; the thin dotted curves assume $\zeta \propto m_h^{2/3}$ with these same $\mmin$.}
\label{fig:qz}
\end{figure}

Figure~\ref{fig:qz}\emph{a} shows ionization histories with $\bar{C}=0,\,1,$ and $3$ (dotted, solid, and dashed curves, respectively).  In all cases, we use the \citet{hopkins07} luminosity functions to compute the emissivity, as well as an ultraviolet spectral index $\alpha=1.6$.  The ordinate, $Q(z)$, is simply the time integral of equation~(\ref{eq:xevol}); it is the total number of ionizing photons produced per helium atom, minus the number of recombinations per atom, and can exceed unity once helium is fully reionized.\footnote{In a realistic model, $\bar{C}$ must increase rapidly as $\bxion \rightarrow 1$, because then dense pockets of gas must begin to be ionized (FO05).  This will of course keep $\bxion \le 1$, but we do not attempt to enforce this limit.  Thus, the very final stages of reionization may be more extended than depicted in our models.}

Given the observational evidence that helium was reionized at $z \sim 3$, it is reassuring to find that the \citet{hopkins07} luminosity function yields $Q=1$ at $z \sim 3$--$3.6$ in these models.  There are a number of reasons why there may be a delay; for example, some of the ionizing photons have such high energies that they will not be absorbed in the diffuse IGM, our assumed spectrum may be too soft, and the clumping factor increases throughout reionization, especially when $\bxion \approx 1$.
Another uncertainty is in the faint-end slope, which more recent data shows to be somewhat steeper than assumed here \citep{fontanot07, bongiorno07, siana08}.  If we fix all the other parameters in the luminosity function, changing the faint end slope to have a constant shape $\propto L^{0.7}$ increases the total luminosity density by $\sim 50\%,\,40\%,$ and $35\%$, respectively, at $z=3,\,4$, and $5$ (the difference occurs because the bright end slope flattens toward high redshifts).  Thus a steeper faint end will mean more photons, but it will not substantially change the \emph{shape} of the curves in Figure~\ref{fig:qz}.\footnote{Another way to address this question would be to use the ``Bright" fit of \citet{hopkins07}, which fixes the faint end slope at $L^{0.6}$.  This also shows little effect on the shape of $Q(z)$.}

Figure~\ref{fig:qz}\emph{b} shows reionization histories normalized so that $Q(z=3)=1$ (i.e., reionization completes at $z_{\rm He}=3$).  The thick curves, which nearly overlap, correspond to the histories in Figure~\ref{fig:qz}\emph{b}.  Interestingly, the shape is nearly independent of the recombination rate, so long as $\bar{C}$ is constant and within this range.  If we rewrite equation~(\ref{eq:xevol}) as $\deriv \bxion/\deriv z$, the sink term goes like $\sim (1+z)^{1/2} \bxion(z)$; a fit to our results yields $\bxion(z) \sim (1+z)^{-6}$, much faster than the density and clumping evolution.  Thus, to a good approximation the recombination rate simply tracks the rapidly-evolving ionized fraction and the shape of $Q(z)/Q(z=3)$ is nearly invariant.

\subsection{A Halo-based Interpretation?} \label{halo}

The empirical model described above gives our best handle on the history of $\bxion(z)$, but it does not provide enough information for the ionized bubble models that we examine next.  These require a relation between the ionizing sources and the underlying large-scale density field.  The clearest way to make this connection is to assign the quasars to dark matter halos.  Although it is beyond the scope of this paper to construct a detailed model of quasar hosts, in this section we will use some general arguments to elucidate some aspects of their relationship.

The thin dot-dashed curves in Figure~\ref{fig:qz}\emph{b} use a different form for the emissivity, in which the quasar emissivity is assumed to trace $\fcoll$, the fraction of matter in collapsed halos.  Thus the first term on the right hand side of equation~(\ref{eq:xevol}) becomes
\bq
\zeta_{\rm He} {\deriv \fcoll \over \deriv t},
\eq
where $\zeta_{\rm He}$ is the number of helium-ionizing photons produced per helium atom in these collapsed halos.  For the minimum mass of a collapsed halo, we will take $\mmin=c_m m_i$, where $m_i$ corresponds to a virial temperature $T_{\rm vir}=2 \times 10^5 \kel$, the approximate collapse threshold for halos in an ionized medium, and $c_m$ is a redshift-independent constant accounting for the possibility that quasars only reside in massive halos.\footnote{Although in principle $c_m$ could be a function of redshift, we will assume for simplicity that it is a constant.}  For consistency with the excursion set model outlined below, we compute $\fcoll$ using the \citet{press74} mass function.  Although the \citet{sheth99} mass function does match simulations better, the relation between halos and quasars is so uncertain that we regard the simpler form as reasonable (and more amenable to calculations).  In general, the abundance of high-mass objects is larger in the \citet{sheth99} mass function, so for a given ionizing efficiency reionization would complete earlier. 

For quasar-like sources, the efficiency can be parameterized as
\bq
\zeta_{\rm He} = 7.3 f_{\rm esc} \left( {f_{>54.4} \over 0.14} {150 \eV \over \VEV{E_{\rm ion}}} {\epsilon \over 0.05} {f_{\rm BH} \over 10^{-5}} \right),
\label{eq:zetadefn}
\eq
where $f_{\rm esc}$ is the fraction of helium-ionizing photons that escape the host galaxy of the quasar, $f_{>54.4}$ is the fraction of the quasar output emitted with $E>54.4 \eV$, $\VEV{E_{\rm ion}}$ is the mean energy of these helium-ionizing photons, $\epsilon$ is the radiative efficiency of the quasar (so the total energy output is $\epsilon m_{\rm BH} c^2$, with $m_{\rm BH}$ the black hole mass), and $f_{\rm BH} = m_{\rm BH}/m_h$ is the fraction of the halo mass $m_h$ inside the central black hole.  To set these fiducial parameters, we have used the optical and ultraviolet mean quasar spectra of \citet{vandenberk01} and \citet{telfer02}, respectively, and the local relation between $m_{\rm BH}$ and $m_h$ from \citet{ferrarese02}.

The upper and lower dot-dashed curves in Figure~\ref{fig:qz}\emph{b} show models with $\mmin=m_i$ and $10 m_i$, respectively, and $\bar{C}=0$.  Both are again normalized to complete reionization at $z=3$, which requires $\zeta_{\rm He}=10$ and $33$ in the two cases. Interestingly, these evolve significantly more slowly than the \citet{hopkins07} model, implying either that the black hole mass-halo mass relation must evolve with redshift or that quasars reside in more massive halos than our simple halo model assumes (although, as we have emphasized, the faint end at $z \ga 4$ is quite uncertain, so the magnitude of the discrepancy is still unknown).  A variety of other evidence points to a characteristic host mass of $\sim 3 \times 10^{12}$--$10^{13} \Msun$, suggesting that $c_m \ga 10$ is appropriate (e.g., Figure~2 of \citealt{lidz06-qsoclust}).  

To develop some more intuition about the quasar-halo relation, it is useful to compare the quasar number density to that of halos.  At $z=3$, the number density of quasars with $L>10^{11.2} \Lsun$ (roughly those dominating the photon budget, according to Figure~\ref{fig:qlf}\emph{b}) is $n_{q} \sim 4 \times 10^{-6} \Mpcden$.  Because each quasar is active for only a short time, the total number density of quasar hosts is $n_{\rm host} \sim n_{q} [H(z) t_{\rm QSO}]^{-1} \sim 10^{-3} t_7^{-1} \Mpcden$.  Depending on the assumed halo mass function, this requires $\mmin \sim 4$--$6 \times 10^{11} \Msun \sim 10 m_i$, roughly consistent with the reionization histories shown above.

Of course, there are many lower luminosity quasars as well, although according to the empirical quasar luminosity function (Fig.~\ref{fig:qlf}\emph{b}) they contribute a relatively small fraction of ionizing photons.  \citet{hopkins05a, hopkins05b, hopkins05c} have argued that lower-luminosity quasars are simply long, but relatively dim, phases in the same high-mass black holes that form high-luminosity quasars.  
Our fiducial model, with a high value of $\mmin$, is meant to mimic the qualitative properties of this kind of quasar-host prescription.

A second check is whether this simple picture reproduces the observed clustering of high-redshift quasars.  \citet{shen07} found that $\xi(r) \approx (r/r_0)^{-\gamma}$ with $\gamma = 2.0 \pm 0.3$ and $r_0=15.2 \pm 2.7 h^{-1} \Mpc$ at $z>2.9$, over the range $4 < (r/h^{-1} \Mpc) < 150$, for the relatively luminous quasars observed with the Sloan Digital Sky Survey, with some evidence that the correlation length increases with redshift out to $z=4$ even more strongly than one would expect from a volume-limited sample.   In our cosmology, this suggests a mean bias $\sim 5.5$ for the quasar halos, again corresponding to $\mmin \sim 6 \times 10^{11} \Msun$.  

Despite this reassuring consistency, there is obviously a great deal more work to be done to model the relationship between quasars and their host halos.  In particular, such a simple-minded picture does \emph{not} quantitatively reproduce the quasar luminosity function as a function of redshift.  In particular, simple one-to-one prescriptions tend to overpredict the abundance of faint quasars and produce a luminosity function that is too shallow at the bright end. However, it is reasonable to expect that a more sophisticated flavor of this class of models could:  various modifications to the black hole-host halo relation allow one to simultaneously fit the luminosity function and clustering as a function of redshift \citep{wyithe02-lf,wyithe05-qsoclust, hopkins07-fp,hopkins07-clust}.  Next we illustrate some of the modifications required in such models.  

The \citet{hopkins07-frame} model remedies this problem by smoothly reducing the probability that both large and small halos host quasars, so that most quasar hosts are near a single characteristic mass; however, the width of this distribution is not well-constrained (e.g., \citealt{lidz06-qsoclust}).  The motivation is that major mergers of large, gas-rich galaxies drive the formation of quasars, and these are most common in hosts just above a characteristic mass.  \citet{wyithe02-lf} use a similar merger picture to steepen the bright end of the luminosity function.  They also  do not impose a minimum host mass and so produce a steeper faint end as well, more consistent with some of the recent data.  

Another problem is that, at $z=3$, $\fcoll(>10m_i) \sim 0.03$, requiring a significantly larger ionizing efficiency than our fiducial choice in equation~(\ref{eq:zetadefn}).  Fortunately, there are several ways to alleviate this discrepancy.  For example, many merger-driven models for quasar formation predict an increasing black hole mass fraction with redshift, ranging from $m_{\rm BH}/m_h \propto (1+z)^{0.5}$ \citep{hopkins07-fp} to $m_{\rm BH}/m_h \propto (1+z)^{5/2}$ \citep{wyithe02-lf}.  (Both models appeal to quasar feedback to set the final relation but differ in how they relate the parameters of the central region surrounding the black hole to the global properties of the host halos).  On the other hand, these scenarios would slightly worsen the discrepancy between the more slowly-evolving $\bxion(z)$ in these $\fcoll$-based models and the faster-evolving empirical calculation.

Another possibility is that quasar luminosity scales more steeply than linearly with halo mass.  \citet{wyithe02-lf} argue that $m_{\rm BH} \propto m_h^{5/3}$, over and above the redshift dependence.  Because more massive halos form later, this will also help to accelerate the evolution in $\bxion$.  The thin dotted curves in Figure~\ref{fig:qz}\emph{b} show the difference:  they assume $\zeta \propto m_h^{2/3}$, with $c_m=1$ and $10$ for the upper and lower curves.  The acceleration is clear, and by weighting more massive galaxies more heavily we can indeed reach a shape very close to the empirical one.  However, \citet{hopkins07-fp} predict a much shallower relation with between black hole and halo mass (although still slightly superlinear).  The differing evolution of black hole mass with redshift in these models offsets the different black hole-halo mass relationships, allowing both to reproduce observations at the bright end.  Both models provide at least approximate fits to the observed clustering \citep{wyithe05-qsoclust, hopkins07-clust}.

Given that these more successful models appeal to mergers, is a simple halo-based interpretation really sufficient for our purposes?  Instead, one might need to include only those halos that are growing rapidly \citep{cohn07}.  Such a change would induce three changes in our $\fcoll$-based model.  First, because major mergers are rare events, it would increase the amount of ``stochasticity" in the quasar hosts relative to the halos.  We will see in \S \ref{stochastic} how to describe this important aspect without appealing to either halos or mergers as hosts, so this aspect will not affect our results.  Second, $Q(z)$ could differ, since the halo merger rate and collapse rate evolve differently with time. Third, the differing clustering of mergers could substantially modify the bubble distribution when it is driven by clustering. We examine these two effects below.  

\begin{figure}
\plotone{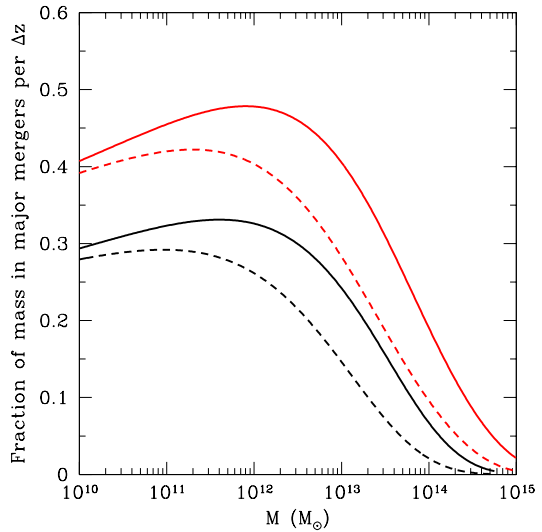}
\caption{Fraction of original mass accreted by halos in mergers per unit redshift.  The solid and dashed curves are at $z=3$ and 4, respectively.  All curves assume $m_u=M$; the upper and lower sets take $m_l=M/10$ and $M/4$, respectively.}
\label{fig:merger_rate}
\end{figure}

The second effect is illustrated in Figure~\ref{fig:merger_rate}, which shows the rate at which galaxies accrete mass in mergers according to the extended Press-Schechter model \citep{lacey93}.  More precisely, the curves show 
\bq
{1 \over M} \int_{m_l}^{m_u} \deriv m \, m {\deriv^2 p(M,m,z) \over \deriv m \, \deriv z},
\label{eq:merger_rate}
\eq
where $\deriv^2 p(M,m,z)/\deriv m \, \deriv z$ is the rate at which halos of mass $m$ merge with a given halo of mass $M$ per unit redshift.  We assume $M>m$ and integrate over all mergers with $m_l < m < m_u$.  We always take $m_u=M$ (in order to avoid double-counting mergers); the upper and lower sets of curves take $m_l/M=0.1$ and $0.25$, respectively, showing two different possible thresholds for triggering a quasar phase. 

Figure~\ref{fig:merger_rate} shows that the merger rate for halos of \emph{any} mass increases with decreasing redshift, independent of the threshold.  This means that merger models will \emph{steepen} $Q(z)$ relative to our $\fcoll$ models, easing the tension with the empirical model.  

The third effect is also illustrated in Figure~\ref{fig:merger_rate}:  the variation in the accreted mass fraction (normalized to the initial mass $M$) is surprisingly independent of $M$, except at masses well above the exponential cutoff in the mass function.  For example, it varies by only $\sim 5\%$ over the entire $10^{10}$--$2 \times 10^{13} \Msun$ range at $z=3$.  Thus the range of halos that would contribute ionizing photons in a merger model is almost the same as in our $\fcoll$ models.  The slight increase with mass for relatively small halos is much smaller than that in our $\zeta_{\rm He} \propto m_h^{2/3}$ model, so such a picture would be bracketed by our two sets of models.  The major difference with our fiducial $\fcoll$ prescription would be to down-weight the most massive halos; this is why both the \citet{hopkins07-frame} and \citet{wyithe02-lf} steepen at the bright end.  This is not too important to our picture, because the photon budget is dominated by halos near $\mmin$.  

Note that we use the extended Press-Schechter merger rates here;  these have recently been shown to provide a relatively poor match to merger rates in the Millenium simulations \citep{fakhouri08}. However, \citet{fakhouri08} find the same qualitative behavior in the simulations that we do; in particular, they show that the mean merger rate per halo depends very weakly on the halo mass ($\propto M^{0.08}$).  Extended Press-Schechter overestimates the rates for equal mass mergers of $10^{12} \Msun$ halos by up to $\sim 80\%$ at these redshifts \citep{fakhouri08, zhang08}.  However, the error decreases to $\sim 40\%$ for 10:1 mergers, and continues to decrease slowly for smaller mass halos.  Scaling the merger rates in this way does not affect our conclusions, because the dependence on mass is still gentle.

In summary, matching $\bxion(z)$ suggests that quasars appear only in massive halos (with $c_m \ga 10$) or that their luminosity scales steeply with host mass, although a detailed fit to the luminosity function requires much more machinery than we have introduced here.  Given our uncertainties in the high-$z$ quasar luminosity function and (especially) in the physics underlying the black hole-host relation, we will not attempt to construct a more detailed or accurate model here; we simply note that the \citet{wyithe02-lf} and \citet{hopkins07-fp} models are specific, and much more well-developed, cases of our more general picture.  Instead we will examine how this relation will affect the morphology of helium reionization.  For instance, we explore the case where $\zeta_{\rm He} \propto m_{h}^{2/3}$, which reproduces the main features of the \citet{wyithe02-lf} model, and a case where $c_m=10$, roughly reproducing the characteristic mass of the \citet{hopkins07-frame} model (although we do not directly appeal to mergers, as the original models did).  Our qualitative conclusions will apply to any history, although some quantitative aspects will depend on the details of the relation.

\subsection{A Stellar Contribution?} \label{stellar}

While it is generally agreed that quasars drive helium reionization, it is worth noting that stellar systems \emph{can} produce photons above the \ion{He}{2} ionization threshold.  In fact, the composite spectrum of $z \sim 3$ Lyman-break galaxies constructed by \citet{shapley03} has a reasonably strong \ion{He}{2} $\lambda 1640$ recombination line, even though it has little (if any) AGN contamination.  The origin of this line is unclear.  It is quite broad (FWHM$\sim 1500 \kms$) and so is most likely atmospheric instead of nebular.  However, \citet{shapley03} showed that known stellar populations, such as Wolf-Rayet stars, cannot simultaneously reproduce the \ion{He}{2} line strength and the metallicity implied by other spectral features.  \citet{jimenez06} have argued that the line may originate from massive Population III stars that are able to form because of inefficient metal mixing.  There have also been suggestions that the \ion{He}{2} $\lambda 1640$ line emission could arise collisionally \citep{yang06}. 

In some of these scenarios, the \ion{He}{2} $\lambda 1640$ luminosity may be accompanied by \ion{He}{2} ionizing photons.  This is the Balmer-$\alpha$ transition, so it must result from either recombinations following ionizations or collisional excitation to the $n=3$ state.  In the latter case, helium line cooling radiation peaks at $T \sim10^5 \kel$, where collisional ionization is weak; the gas tends to be optically thick to 4 Ryd photons unless $T > 10^6 \kel$ \citep{miniati04}, so there may not be much ionizing flux.  But if the origin is radiative, or if the gas is hot, these photons would be accompanied by a substantial ionizing flux.  We now show that, if a substantial fraction of these photons escape into the IGM, they may play a significant role in helium reionization.

Rather than work within the context of a specific model, we will instead simply use the \citet{shapley03} spectrum to associate the \ion{He}{2} $\lambda 1640$ luminosity, and hence the production rate of ionizing photons, to the star formation rate.  (We could instead associate the luminosity to the total stellar mass, but it seems more likely that these high-energy photons are produced by young stars.)  First, we note that the flux density at this line is $f_\nu \approx 0.35 \, \mu$Jy, and the line has an equivalent width $W_\lambda \sim 2$ \AA \ \citep{jimenez06}.  Assuming that both kinds of photons result from radiative recombinations, the line luminosity is related to the ionizing photon production rate $Q_i$ via \citep{schaerer03}
\bq
L = c_1 (1 - f_{\rm esc}) Q_i,
\label{eq:schaerer}
\eq
where $f_{\rm esc}$ is the escape fraction of the photons and $c_1=5.67 \, (6.40) \times 10^{-12} \erg$ for gas at $30,000 \, (10,000) \kel$; we will assume the lower temperature for our estimates.  From these, together with the average star formation rate in the \citet{shapley03} sample ($36 \sfr$), we find that the number of helium-ionizing photons produced per helium atom in stars, $\eta$, is
\bq
\eta \sim {70 \over 1 - f_{\rm esc}} \left( {W_\lambda \over 2 \mbox{ \AA}} \right).
\label{eq:etacalc}
\eq
The number of photons that escape into the IGM, per helium atom in the universe, is then
\bq
Q_\star(z) = \zeta_{\rm He}^\star \fcoll = \eta f_{\rm esc} f_\star \fcoll \sim 2 {f_{\rm esc} \over 1 - f_{\rm esc}} \left( {\fcoll f_\star \over 0.03} \right),
\eq
where $\zeta_{\rm He}^\star$ is a measure of the ionizing efficiency per helium atom in galaxies and $f_\star$ is the overall star formation efficiency.

To estimate the factor $f_\star \fcoll$, we note that $\sim 10\%$ of all baryons are in stars at the present day \citep{fukugita04}, and according to recent measurements $\sim 1/3$ of these formed by $z=3$ (e.g., see Fig.~1 of \citealt{panter07}, and references therein).  Thus at $z=3$, $f_\star \fcoll \sim 0.03$.  Comparing to Figure~\ref{fig:qz}, this stellar component will be competitive with quasars if $f_{\rm esc} \ga 0.5$.

The escape fraction is difficult to estimate, but it is possible (at least in principle) that it is larger than that for hydrogen (which is known to be at most a few percent; \citealt{shapley06}). For instance, the large line widths of observed HeII recombination lines ($\sim 1500 \, {\rm km \, s^{-1}}$) imply that He ionizing photons may arise from a different (or subset) stellar population -- likely associated with stellar winds and outflows -- from the general OB stellar population that produces hydrogen ionizing photons. Such a population with fast winds could possibly have a large fraction of ``clear" sightlines.  

Thus we find that, even with the substantial helium recombination lines in moderate-redshift galaxies, stellar systems will probably not dominate the helium ionization budget, unless a large fraction ($\sim 0.5$) of those photons can escape into the IGM.  On the other hand, they may not be a completely negligible component, especially since the resulting photons will probably be produced quite close to the helium ionization edge (and thus interact in the nearby IGM) and because the stellar component evolves less rapidly than the quasar component (so may be more important at high redshifts).  While it is certainly speculative to attribute helium reionization to this galactic emission, it could significantly impact the large observed hardness fluctuations in the radiation field on small ($\sim 1 \Mpc$) scales after reionization \citep{shull04}.

\section{Stochastic Reionization}
\label{stochastic}

We now turn to modeling the morphology of helium reionization.  We will begin by examining the limit in which the ionizing sources are relatively rare and bright, so that random fluctuations in their distribution determine the ionization morphology.  Of course, we do expect that quasars trace the underlying density field, so their locations are not truly random.  But so long as the number of sources per discrete ionized 
bubble\footnote{See \S \ref{blur} below for a discussion of our discrete bubble approximation.} is less than a few, random fluctuations in that number will dominate over clustering in setting the typical bubble 
size.\footnote{Note that here we refer to the number of sources in a bubble, integrated over all time, not just the active quasars at any given instant.  Most bubbles grow from multiple sources but will still have only one \emph{active} quasar at any given time, because of their short duty cycle.  We thus implicitly assume that the bubbles do not significantly recombine between the episodic ionizations (but see \S \ref{fossils}).  We also assume that a single halo does not host repeated quasar generations; in that case the bubbles would grow monotonically around these unusual halos and reach much larger sizes than estimated here.}   This is different from hydrogen reionization, where the sources are small and numerous, so fluctuations can typically be ignored \citep{furl05-charsize}.

For a simple toy model, we begin by assuming that all the ionizing sources have an identical luminosity and that internal absorption within each ionized bubble is negligible.  The first assumption is clearly not correct in detail, but at $z=3$, $\sim 50\%$ of the ionizing photons are produced by quasars with $L \sim 0.3$--$3 L_\star$, so it is not a terrible approximation.\footnote{The fraction falls to $\sim 35\%$ with the steeper faint end slope of recent data \citep{fontanot07, bongiorno07, siana08}.  Obviously the distribution of luminosities will be more important in this case.} We will examine the second assumption more closely below (see \S \ref{recomb}).  These two assumptions then imply that the ionized bubbles are built of units with fixed volume $V_i$.

We will further assume that the ionizing sources are Poisson-distributed with number density $n_{\rm src}$.  \citet{sheth-lemson99} and \citet{casas02}  found that the variance of halo counts in simulations of the $z=0$ universe is nearly Poissonian in regions ranging from voids to moderate overdensities,  with the discrepancy relative to Poisson no more than a factor of two.  Let us select regions of the universe with a smoothing window $V_i$; we can compute the total ionized fraction by counting the smoothing windows that actually contain sources and multiplying by the number of sources within each such clump:
\bqa
\bxion & = & P(1|V_i) + 2 P(2|V_i) + 3 P(3|V_i) + ... \nonumber \\
& = & \sum_{k=1}^\infty k {\bxion^k \over k!} e^{-\bxion}.
\label{eq:pone}
\eqa
Here $P(k|V_i)$ is the Poisson probability of finding $k$ sources in each window; the prefactor $k$ in the second line accounts for the extra volume ionized by each of these clumps,\footnote{In other words, two nearby sources still ionize a volume 2$V_i$ because of photon conservation.} and we have used $n_{\rm src} V_i = \bxion$.  (It is easy to verify that we recover the correct ionized fraction using the power series expansion of $e^{-\bxion}$.)

Similarly, we can compute the fraction of space inside ionized bubbles containing \emph{at least} two quasars by smoothing over windows $2 V_i$ and including only those regions that are completely ionized.  The same procedure works for bubbles with $V>NV_i$, which must surround networks of $N$ 
sources.\footnote{One can imagine that, when $\bxion$ is sufficiently large, this smoothing procedure will not capture some configurations.  However, we will see below that the stochastic stage does not apply when $\bxion \ga 0.5$ anyway.}  Then
\bqa
\bxion(\ge N) & = & {N \over N} P(N|V_i) + {N+1 \over N} P(N+1|V_i) + ... \nonumber \\
& = & {1 \over N} \sum_{k=N}^\infty {(N \bxion)^k \over (k-1)!} e^{-N \bxion} \nonumber \\
& = & \bxion \frac{\Gamma(N) - (N-1) \Gamma(N-1,N \bxion)}{\Gamma(N)},
\label{eq:pgn}
\eqa
where the $1/N$ prefactor in each term appears because we are counting volumes in units of $N V_i$.  The fraction of space filled by bubbles with precisely $N$ sources is then simply
\bq
\bxion(N) = \bxion(\ge N) - \bxion(\ge N+1).
\label{eq:pn}
\eq

\begin{figure}
\plotone{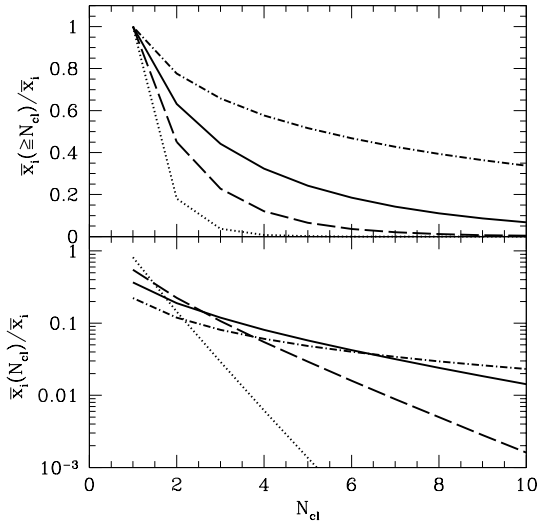}
\caption{Fraction of the ionized volume filled by bubbles with at least $N_{\rm cl}$ sources (top panel) and with exactly $N_{\rm cl}$ sources (bottom panel), assuming that the they are Poisson-distributed.  The dotted, long-dashed, solid, and dot-dashed curves assume $\bxion=0.1,\,0.3,\,0.5$, and 0.7, respectively.}
\label{fig:poisson}
\end{figure}

We show this distribution for several choices of $\bxion$ in Figure~\ref{fig:poisson}.  The top and bottom panels show the cumulative and differential versions, respectively.  Note that we normalize to the total $\bxion$ in each case, so the curves give the fractional contribution of each $N_{\rm cl}$ to the total ionized volume.  Obviously, the fraction inside small bubbles decreases throughout reionization, while the number of clumps grows fairly rapidly.  We find that $\approx 0.17\%,\,6.5\%,$ and $24\%$ of the ionized volume is filled by bubbles with at least five sources when $\bxion=0.1,\,0.3,$ and $0.5$, respectively.

Interestingly, for these randomly distributed sources there is no characteristic scale to the bubble size distribution; instead we find that the distributions approach power laws in the large $N$ limit, with 
\bq
{\bxion(N+1) \over \bxion(N)} \approx 2 \bxion (1- \bxion/2),
\label{eq:poisson-plaw}
\eq
to a maximum absolute deviation of $\sim 0.08$ when $\bxion \sim 0.35$.  We will see that this ``scale-free" behavior is (in principle) different from reionization by clustered sources, which imprints a well-defined characteristic scale on the bubble distribution.  However, in practice the bubbles with one (or at most a few) ionizing sources dominate the distribution, so there will appear to be a characteristic scale of this magnitude in any realistic observed sample.  This apparent scale is, however, illusory in another sense:  we have ignored the luminosity distribution of sources.  In actuality, the size distribution will be broader as it must be averaged over the quasar luminosity function (though it will remain reasonably narrow because $R_i \propto L^{1/3}$).  This is a straightforward exercise for isolated quasars (see Fig.~\ref{fig:qlf}), but it is difficult to include both a luminosity distribution and Poisson fluctuations, and we defer it to future, simulation-based work.

Note that this simple model for stochastic bubbles does not make any assumptions about the quasar sources; it is based purely on the observed luminosity function.  Thus it applies equally well to the halo-based models we appeal to next and to merger-driven models \citep{wyithe02-lf, hopkins07-frame}.  

\begin{figure*}
\plottwo{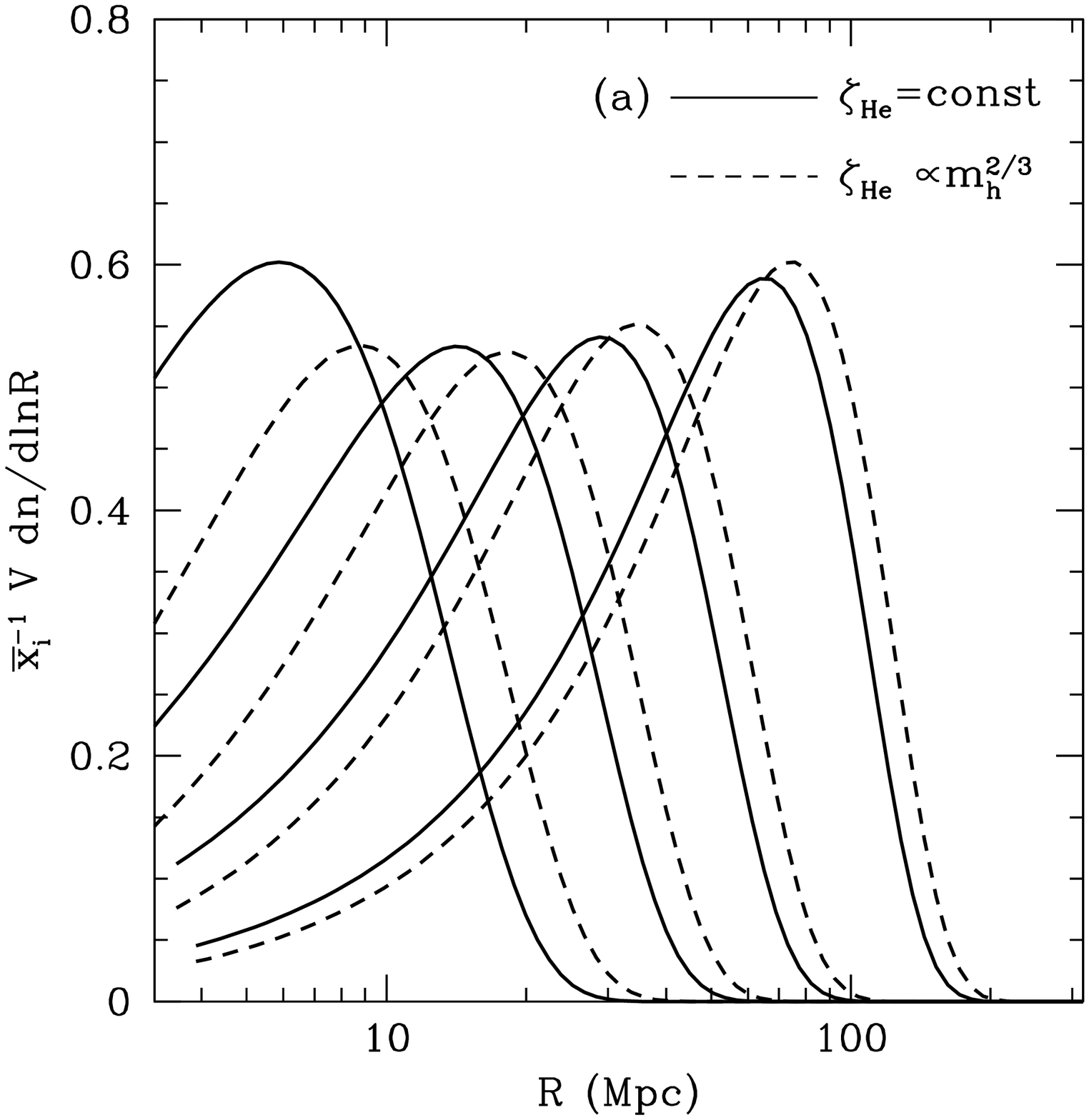}{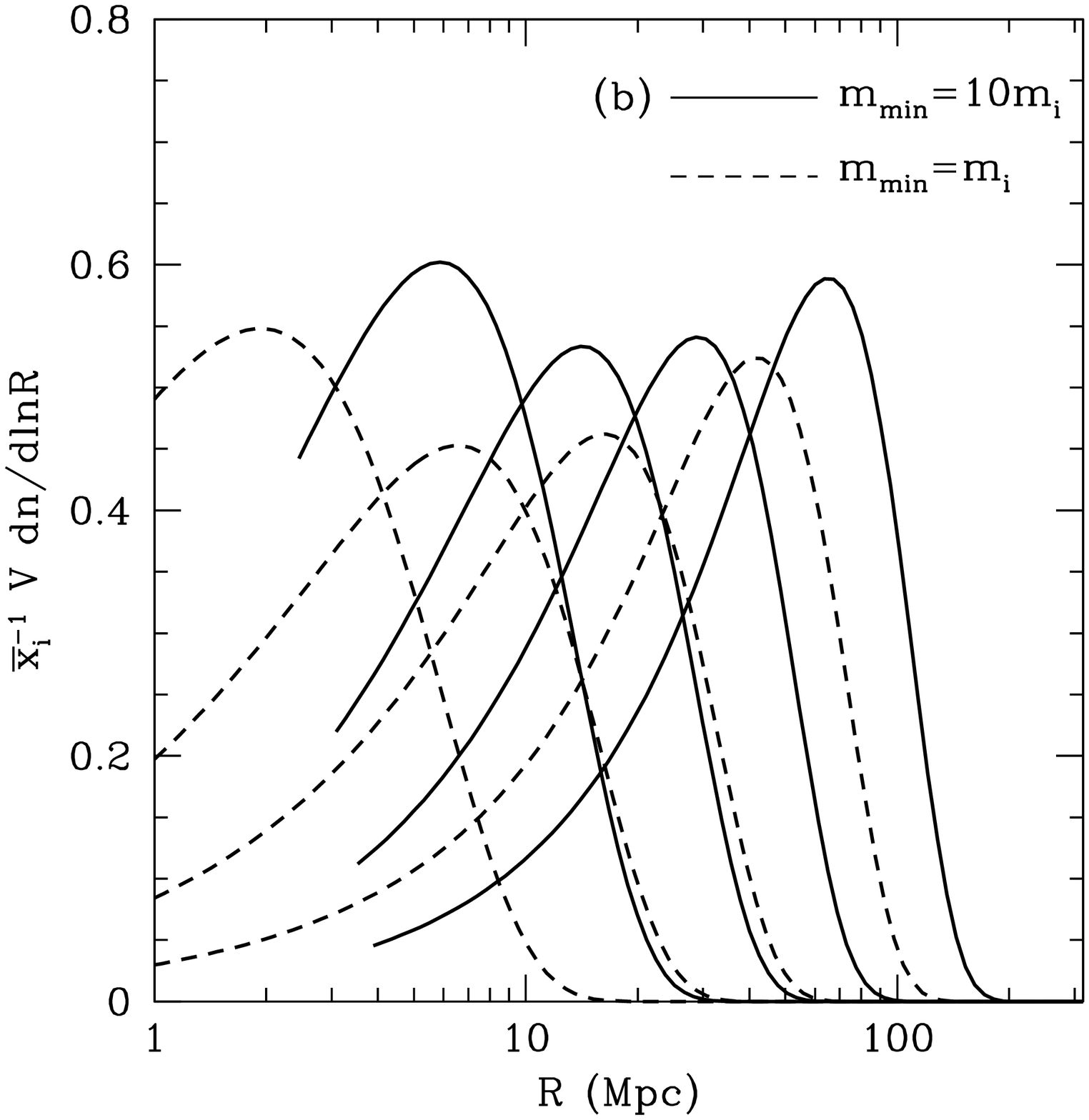}
\caption{Bubble size distributions at $z=4$.  The solid curves in both panels assume $\zeta_{\rm He}$ is independent of halo mass and take $\mmin=10 m_i$.  In \emph{(a)}, the dashed curves take  $\mmin=10 m_i$ with $\zeta_{\rm He} \propto m_h^{2/3}$.  In \emph{(b)}, the dashed curves take $\mmin=m_i$ with a constant ionizing efficiency.  In each set, the curves have $\bxion=0.2,\,0.4,\,0.6$, and $0.8$, from left to right.}
\label{fig:nbub}
\end{figure*}

\section{Ionized Bubbles from Clustered Sources}
\label{bub}

We will now examine the bubble morphology generated by clustered ionizing sources.  A simple way to see that clustering must be important is via the measured quasar correlation function.  Since $r_0=(15.2 \pm 2.7) h^{-1} \Mpc$ for luminous quasars \citep{shen07}, the correlation length is comparable to the ionized bubble radius for an $L_{*}$ quasar, implying that clustering should certainly help large bubbles to build more quickly. 
Moreover, the clustering of halos near quasars certainly affects the local ionizing background during hydrogen reionization \citep{yu05, alvarez07, lidz07}, and it is reasonable to expect the same during helium reionization.

For a more quantitative approach, we begin with the simple ``photon-counting" model of FZH04.  The basic idea is to compare the number of helium-ionizing photons produced by the sources within a patch of the IGM to the number of helium atoms in that patch; wherever the former is greater, we have an ionized bubble.  To begin, we assume a spatially homogeneous recombination rate, with $\bar{N}_{\rm rec}$ the mean number of recombinations per ionized helium atom in the universe (but see \S \ref{recomb}).  If we further assume that every gravitationally-bound halo has a constant ionizing efficiency $\zeta_{\rm He}$, defined as in equation~(\ref{eq:zetadefn}), the criterion for a region with total mass $M$ and fractional overdensity $\delta$ to be ionized by sources contained inside it  can then be written\footnote{Note that, by using the cumulative number of recombinations per \emph{ionized} atom, we are implicitly assuming that the universe consists only of fully ionized and fully neutral regions; see \S \ref{blur} below.}
\bq
\zeta_{\rm He} \fcoll(\mmin,\delta | M) > (1 + \bar{N}_{\rm rec}),
\label{eq:zeta1}
\eq
where $\fcoll(\mmin,\delta | M)$ is the collapse fraction in this region.

This condition~(\ref{eq:zeta1}) can easily be modified to allow the ionizing efficiency to vary across different halos \citep{furl05-charsize}.  If, for example, we allow $\zeta_{\rm He} = \zeta_{\rm He}(m_h)$, it can be written $\bar{\zeta}_{\rm He} \fcoll>1 + \bar{N}_{\rm rec}$, where
\bq
\bar{\zeta}_{\rm He} = {1 \over \fcoll} \int_{\mmin}^\infty dm_h \, \zeta_{\rm He}(m_h) m_h n_h(m_h),
\eq 
and $n_h(m_h)$ is the dark matter halo mass function.  

We will use the \citet{press74} model for $n_h(m_h)$, although other mass functions give nearly identical results when normalized to a constant average ionized fraction $\bxion$, because our model depends only on the variation of the mass function with the large-scale overdensity $\delta$, which is more or less the same for other mass functions \citep{furl05-charsize}.  
Also, as discussed in \S \ref{halo}, more detailed models appeal to mergers rather than halos to host quasars.  \citet{cohn07} presented a model, in the same spirit as FZH04, that follows hydrogen reionization if sources are driven by mergers.  In our case, Figure~\ref{fig:merger_rate} shows that the merger rate is roughly independent of halo mass (except for extremely massive halos).  We will see below that it is only the (luminosity-weighted) clustering that matters for our model, so this implies that basing the source population on halos and mergers will be nearly equivalent.  Thus, we will defer a more detailed, merger-based treatment to the future and focus our initial model on the halo population, as in FZH04.

FZH04 used condition~(\ref{eq:zeta1}), together with the excursion set formalism \citep{bond91,lacey93} to write the mass function of ionized bubbles during reionization.  In essence, for a region of mass $M$ the condition $\zeta_{\rm He} \fcoll(\delta | M)=(1 + \bar{N}_{\rm rec})$ (considered as a function of the overdensity $\delta$) replaces the usual spherical collapse criterion $\delta_c=1.69$ for halo formation.
This assigns each excursion-set trajectory to the largest ionized bubble of which it is a part (thereby implicitly including all of the relevant neighbors).  In order to write the solution analytically, we replace the exact criterion for $\delta$ with a linear fit in $\sigma^2(m)$, the smoothed linear-theory variance in the density field.  The resulting mass distribution of ionized bubbles is
\bq
n_b(m,z) = \sqrt{\frac{2}{\pi}} \ \frac{\bar{\rho}}{m^2} \ \left|
  \frac{\deriv \ln \sigma}{\deriv \ln m} \right| \ \frac{B_0(z)}{\sigma(m)} \exp \left[ - \frac{B^2(m,z)}{2 \sigma^2(m)} \right],
\label{eq:dndm}
\eq
where the excursion set barrier is $\approx B(m,z) = B_0(z) + B_1(z) \sigma^2(m)$ and $\bar{\rho}$ is the mean comoving mass density.  Note that the mean recombination rate has no effect on the result (at a fixed $\bxion$), because it is completely degenerate with the ionizing efficiency.  (Inhomogeneous recombinations, on the other hand, have a substantial effect; see \S \ref{recomb} below.)

Thus the original FZH04 model is easily portable to the case of helium reionization, requiring only a re-definition of the ionizing efficiency.  Figure~\ref{fig:nbub} shows some example distributions.  For our fiducial model (solid curves), we assume that $\mmin=10 m_i$ and that $\zeta_{\rm He}$ is independent of halo mass, which gives an ionization history close to (but slightly slower than) the \citet{hopkins07} luminosity function (Fig.~\ref{fig:qz}).  Panels \emph{(a)} and \emph{(b)} contrast this with models with $\zeta_{\rm He} \propto m_h^{2/3}$ and with $\mmin = m_i$, respectively.

For our fiducial model, the distributions peak at $R \sim 4,\, 12,\, 30,$ and $60 \Mpc$ at $\bxion=0.2,\,0.4,\,0.6,$ and $0.8$, respectively.  The bubbles are somewhat larger if $\zeta_{\rm He}$ is an increasing function of mass, and somewhat smaller if lower-mass halos contribute.  This is because the bubble sizes are primarily a function of the luminosity-weighted clustering of the ionizing sources \citep{furl05-charsize}.  

\begin{figure}
\plotone{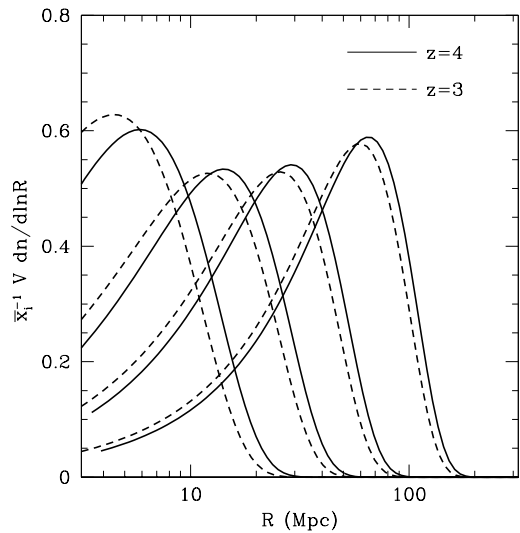}
\caption{Bubble size distributions at $z=4$ (solid curves) and $z=3$ (dashed curves).  Both sets assume our fiducial model.  In each case, the curves have $\bxion=0.2,\,0.4,\,0.6$, and $0.8$, from left to right.}
\label{fig:nbub_z}
\end{figure}

Figure~\ref{fig:nbub_z} contrasts the distributions at $z=4$ and $z=3$.  As with hydrogen reionization, the dependence on redshift is extremely weak:  the bubble sizes are sensitive to the integrated bias of the sources, which does not evolve rapidly in this regime.  They are slightly smaller at the lower redshift, primarily because $m_i$ is a decreasing function of redshift.  

Of course, we have already seen that for reasonable quasar lifetimes the ionized region around a single $L_\star$ quasar at $z=3$ has $R \sim 14 \Mpc$ (they are somewhat smaller at larger redshifts because $L_\star$ decreases); our typical bubbles do not reach such large sizes until $\bxion \sim 0.4$.  It may at first seem that the smaller bubbles are just created by dimmer quasars that produce relatively few ionizing photons.  However, there is actually a fairly fundamental inconsistency here.

The problem occurs because the clustering model assumes that the ionizing bubbles precisely trace the underlying density field and does not self-consistently account for the discrete sources:  in other words, it does not properly account for their stochastic fluctuations, and it forces high-density regions to be ionized even if they may not have a high-mass halo.  This tends to force many small, high-density regions to be ionized, rather than the much larger environments of the real halos.\footnote{In other words, these regions' \emph{average} collapse fractions are large enough to ionize themselves, but not enough to contain an actual halo.  For example, consider a region that is $20\%$ the size of a typical ionized bubble, and that has a $20\%$ chance of containing a halo with $m>\mmin$.  According to the density driven model, every one of these regions would be ionized, even though in reality only one in five would be.  But that one region would also ionize its surroundings, making the total ionized volumes equal in each picture.}  As a result, we must turn to the stochastic model of \S \ref{stochastic} early in reionization.

This can be seen most clearly by examining the characteristic bubble size $R_c$ (defined as the peak of $R \, m \, n_b$) to an analogous quantity in the stochastic reionization model.  As we have seen the size distribution in the latter case is nearly a power law:  there is no well-defined peak.  We therefore compare $R_c$ to the effective number of sources $N_c$ for which $V(>N_c) = \bxion/2$ (i.e., half of the ionized volume is in bubbles smaller than -- or larger than -- this value).  For these purposes, we treat $N_c$ as a continuous, rather than discrete, variable.  We then convert to an effective radius by assuming that $R_c = 14 N_c^{1/3} \Mpc$, where the proportionality constant comes from the peak of the $z=3$ luminosity density in Figure~\ref{fig:qlf}\emph{b} (see eq.~\ref{eq:Vi}) and assumes $f_{\rm abs}=1$, $t_7=1$, and $\bar{x}_{\rm HeII}^u=1$.

\begin{figure}
\plotone{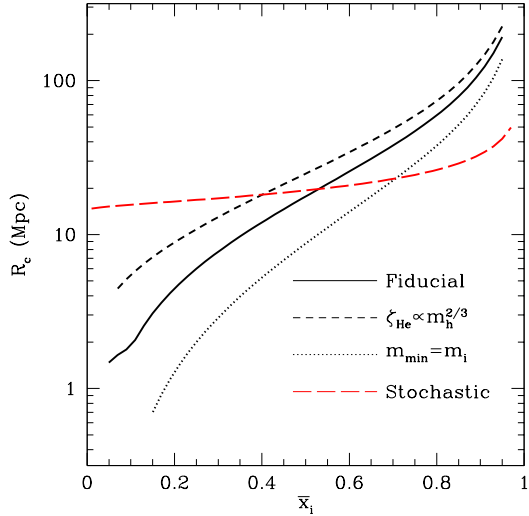}
\caption{Characteristic bubble sizes as a function of $\bxion$ at $z=3$.  The solid curve assumes $\zeta_{\rm He}$ is independent of halo mass and takes $\mmin=10 m_i$.  The dotted curve takes $\mmin=m_i$.  The short-dashed curve takes $\zeta_{\rm He} \propto m_h^{2/3}$.  The long-dashed curve is for our stochastic model (see text for a precise definition).}
\label{fig:rchar}
\end{figure}

Figure~\ref{fig:rchar} shows how these scales evolve in several models.  The solid curve shows our fiducial clustering model, the short-dashed curve assumes $\zeta_{\rm He} \propto m_h^{2/3}$, and the dotted curve assumes $\mmin=m_i$.  The long-dashed curve shows the stochastic model, with $R_c$ defined as above.  As with hydrogen reionization, the characteristic size of the clustering-driven bubbles increases rapidly throughout reionization, surpassing 15 Mpc by $\bxion \sim 0.5$ and eventually reaching $R_c \ga 100 \Mpc$.  At a given ionized fraction, $R_c$ increases with the underlying bias of the sources.

In comparison, $R_c$ defined from the stochastic model increases much more slowly:  by only a factor $\sim 2$ from $\bxion=0.01$ to $\bxion=0.75$.  Thus, only a few sources contribute to each ``typical" bubble, regardless of the ionized fraction.  
The density-driven model predicts much smaller bubbles early on, with only a small fraction of the Universe contained in bubbles above the stochastic scale.  However, as $\bxion$ increases $R_c$ increases much more rapidly for the clustering model than for the stochastic model, eventually overtaking it.  During the latter stages of reionization, the stochastic component will become negligible compared to the typical $\sim 100 \Mpc$ bubbles that emerge; by this point, fluctuations in the clustering networks are no longer important.

Of course, we have built our stochastic model from bubbles with a fixed size, appropriate for $L_\star$ quasars.  This is reasonable with the \citet{hopkins07} luminosity function, which is rather strongly centered around $L_\star$ at high redshifts.  With the steeper faint end slope indicated by more recent data, fainter quasars contribute somewhat more of the luminosity density.  Thus the stochastic distribution will be built of somewhat smaller units, and we would expect clustering to become important somewhat earlier.

This suggests that we can approximately account for the stochastic component by assigning all of the volume nominally in small bubbles to the characteristic scale of the stochastic component.  Of course, that distribution is relatively uninteresting (depending only on the mean number density and luminosity of sources), so it is only once the typical bubble grows beyond the stochastic scale that we can study the astrophysics of the sources via their effects on the bubbles.  For example, we have seen that the bias of the quasar host population affects the bubble sizes.  Alternatively, if there is a significant population of small bubbles early in the reionization process, this raises the interesting possibility that there is an unresolved population (such as miniquasars or galaxies) that contributes to helium reionization. 

\subsection{Bubbles of Ionized Helium?}
\label{blur}

In the previous two sections, we have treated the ionized zones around quasars as having sharp, well-defined edges.  This is necessary for the excursion set model of FZH04, which implicitly assumes that gas is either fully ionized or fully neutral.  It is an excellent assumption for hydrogen reionization; the mean free path of a hydrogen-ionizing photon is $\lambda_{\rm HI} \approx 0.002 ([1+z]/10)^2 (\nu/\nu_{\rm HI})^3 \Mpc$, where $\nu_{\rm HI}$ corresponds to the ionization edge of neutral hydrogen.  The stellar sources generally assumed to be responsible for hydrogen reionization have rather soft spectra, so $\VEV{\nu} \approx \nu_{\rm HI}$.  Thus, once an ionizing photon hits the edge of an ionized bubble, it is quickly absorbed.  The transition regions between ionized and neutral gas have negligible thickness, so the two-phase approximation is an excellent one.

In the case of helium, several factors make this assumption problematic.  First, helium is less common and has a smaller photo-ionization cross-section.  Second, helium reionization occurs at $z \sim 3$, when the universe is much more dilute.  As a result, in a homogeneous medium
\bq
\lambda_{\rm He} = 0.66 \left( {4 \over 1+z} \right)^2 \left( {\nu \over \nu_{\rm HeII}} \right)^3 \Mpc,
\eq
where $h \nu_{\rm HeII} = 54.4 \eV$.  Third, and most importantly, the quasars thought to be responsible for helium reionization have rather hard spectra.  At the mean spectral index, $\alpha = 1.6$, only half the photons have $\nu \la 1.5 \nu_{\rm HeII}$ and hence have mean free paths $\lambda_{\rm He} < \lambda_{1/2} \approx 2.3 [4/(1+z)]^2 \Mpc$.  Thus we expect the typical ``transition region" around each ionized bubble to be  a few comoving Mpc thick.  Fortunately, we have seen that the bubbles exceed this size throughout most of reionization (see eq.~\ref{eq:Vi}), so our two-phase approximation should be reasonable.  We also note that the only published full numerical simulation of reionization \citep{sokasian02} also made the two-phase approximation by assuming that all photons lay at the ionization edge.

The hard photons that do escape to large distances well outside the transition region form a more uniform, low-level ionizing background.  Bubble expansion in a uniformly ionized medium can easily be incorporated into the FZH04 model by changing the ionization criterion to $\zeta_{\rm He} \fcoll > \bar{x}_{\rm HeII}^u (1 + \bar{N}_{\rm rec})$ \citep{furl04-21cmtop}; the net result is to make the bubbles somewhat larger for a given total ionized fraction than they would otherwise be, but the qualitative behavior is unchanged.  

There is one additional complication:  the IGM is, of course, clumpy and not uniform.  We will next describe how this affects the bubble distribution.

\section{Inhomogeneous Recombinations and the Bubble Distribution}
\label{recomb}

To this point, we have ignored the effects of recombinations on the bubble distribution (treating them as a spatially uniform photon sink).  FO05 showed that \emph{inhomogeneous} recombinations become increasingly important during hydrogen reionization as denser and denser regions are ionized.  They constructed a second excursion set barrier that corresponded to the (instantaneous) ionization rate balancing the recombination rate inside the bubbles (with the latter set by the requirement that ionizing photons be able to reach the edge of the bubble, which sets the minimum density of neutral regions via mean free path arguments).  This requirement becomes important in the late stages of reionization and effectively caps the bubble growth; qualitatively, this maximum size corresponds to the mean free path between optically thick Lyman-limit systems in the early universe.  The transition to the recombination-limited regime enforces a smooth match from the highly inhomogeneous bubble-dominated morphology characteristic of reionization to the more uniform radiation background of the post-reionization universe.

The FO05 excursion set barrier implicitly assumes that the rate at which ionizing sources produce photons is steady.  This is an excellent approximation for the small sources expected to dominate the hydrogen era, especially if their star formation timescales are long \citep{furl05-charsize}.  However, helium reionization is very different:  as we have seen, bubbles are typically produced by only a few sources, and the duty cycle is very short:  $t_{\rm QSO} H(z) \sim 0.003 t_7 [(1+z)/4]^{3/2}$.  Thus even the largest bubbles are likely to contain only one (or at most a few) active quasars at any given time.  In these bubbles, the emissivity fluctuates strongly, at zero most of the time but far above the mean value during active phases.  Thus, we really only need to consider IGM recombinations around isolated quasars.

To that end, consider a gas element in the IGM with overdensity $\Delta = \rho/\bar{\rho}$.  In the highly-ionized limit, ionization equilibrium demands that
\bqa
x_i & = & { \alpha n_e \over \Gamma } \nonumber \\
& = & 3.8 \times 10^{-3} \Delta T_4^{-0.7} \left( {10^{-14} \secinv \over \Gamma} \right) \left( {1+z \over 4} \right)^3, 
\label{eq:ioneq}
\eqa
where we have used the case-A recombination rate, $T = 10^4 T_4 \kel$ is the IGM temperature, and the ionization rate is
\bqa
\Gamma & = & (1+z)^2 \int_{\nu_{\rm HeII}}^\infty {L_\nu \over 4 \pi R^2} \, {\sigma_\nu \over h \nu} \deriv \nu.
\label{eq:gammadefn} \\
& = & 1.74 \times 10^{-14} \secinv \left( {L_B \over 10^{12} \Lsun} \right) \left( {10 {\rm Mpc} \over R} \right)^2 \left( {1+z \over 4} \right)^2.
\eqa
Here $\sigma_\nu = 1.91 \times 10^{-18} (\nu_{\rm HeII}/\nu)^3 \cmsq$ is the photoionization cross section for \ion{He}{2}, $R$ is the \emph{comoving} distance from the quasar, and we have used the mean quasar spectrum ($\alpha=1.6$).

\citet{schaye01} has shown that \ion{H}{1} \lya forest absorbers can be accurately modeled by assuming their physical scale to be comparable to the Jeans length $L_J \approx c_s/(G \rho)^{1/2}$.  We will use the same approximation for helium, so that an absorber with overdensity $\Delta_i$ has column density $N_{\rm HeII} \approx L_J x_i n_{\rm He}$.  Then
\bq
N_{\rm HeII} \approx 1.8 \times 10^{15} \Delta^{3/2} T_4^{-0.2} \left( {1+z \over 4} \right)^{9/2} \left( {10^{-14} \secinv \over \Gamma} \right).
\label{eq:nheii}
\eq
A system will become optically thick when $\tau_\nu = \sigma_\nu N_{\rm HeII} = 1$.  At the ionization edge, this requires an overdensity $\Delta_i$
\bq
\Delta_i \approx 64 T_4^{2/15} \left( {L_B \over 10^{12} \Lsun} \right)^{2/3} \left( {{\rm Mpc} \over R} \right)^{4/3} \left( {1+z \over 4} \right)^{-5/3}.  
\label{eq:deltai}
\eq
Note that this definition of $\Delta_i$ uses the photoionization cross section at the ionization edge; with the hard spectra typical of quasars, many of the photons have much higher energies and so can penetrate even denser systems than implied by this simple model (see \S \ref{blur} above).  To account for this, one could define a frequency-dependent $\Delta_i(\nu) = \Delta_i (\nu/\nu_{\rm He})^2$.  On the other hand, many high-energy photons will not even interact at the edge of the bubble but instead join the uniform background.

\begin{figure}
\plotone{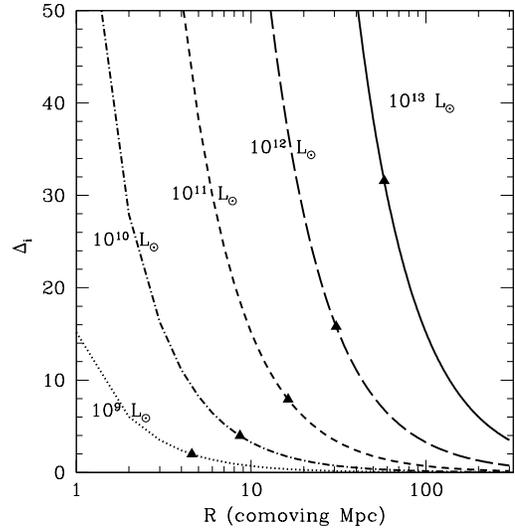}
\caption{Density required to produce an absorption system optically thick to photons at the helium ionization edge, as a function of distance from the central quasar.  The dotted, dot-dashed, short-dashed, long-dashed, and solid curves assume $\log (L_B/{\rm L}_\odot) = 9,\,10,\,11,\,12,$ and 13, respectively.  Along each curve, the solid triangle marks the point at which $R=\lambda(\Delta_i)$ according to the MHR00 model.  We take $z=3$.}
\label{fig:DeltaR}
\end{figure}

Figure~\ref{fig:DeltaR} shows this overdensity $\Delta_i$ as a function of distance for quasars of several different luminosities ($\log [L_B/{\rm L}_\odot] = 9,\,10,\,11,\,12,$ and 13, from top to bottom).  The faintest quasars have $\Delta_i \la 5$ outside of a few Mpc, but typical sources are able to ionize even extremely dense gas up to rather large distances.

To estimate the effects of these optically thick systems on the ionized bubbles, we will use the same fundamental approach as FO05 but take advantage of the much better-known IGM at $z \sim 3$.  In particular, if a region with $\Delta > \Delta_i$ appears between a quasar and the edge of its host bubble, then most of its photons will be consumed overcoming recombinations rather than ionizing new material.  We therefore wish to compare our bubble sizes to the mean free path between IGM patches with $\Delta = \Delta_i$.  To obtain the latter, we need a model for the density structure of the IGM.

\subsection{The IGM Density Distribution} \label{density}

\citet[hereafter MHR00]{miralda00} present a model for the volume-averaged density distribution of the IGM gas, $P_V(\Delta)$, based on a fit to simulations at $z \sim 2$--$4$:
\bq
P_V(\Delta) \, \deriv \Delta = A_0 \Delta^{-\beta} \exp \left[ - \frac{(\Delta^{-2/3} - C_0)^2}{2(2 \delta_0/3)^2} \right] \, \deriv \Delta.
\label{eq:pvd}
\eq
Intuitively, the underlying Gaussian density fluctuations are modified through nonlinear void growth and a power law tail at large $\Delta$.  Here $\delta_0$ essentially represents the variance of density fluctuations smoothed on the Jeans scale for an ionized medium; thus $\delta_0 \propto (1+z)^{-1}$ at  high redshifts.  The power-law exponent $\beta$ determines the behavior at large densities; for isothermal spheres, it is $\beta=2.5$.  We use the fitted values from MHR00 for these parameters; the remaining constants ($A_0$ and $C_0$) can be set by demanding proper mass and volume normalization.

Note that MHR00 base their distribution on a simulation that does not incorporate all the physics of helium reionization, and in particular it does not follow the inhomogeneous heating during that event.  In reality, the clumping will decrease during helium reionization because of photoheating (see, e.g., \citealt{furl07-igmtemp}), which will modify the distribution of densities as well.  This is probably not too important for our purposes, because the IGM takes a fairly long time to adjust to the new Jeans smoothing scale \citep{gnedin98}, and the increase is much more modest than during helium reionization.  

MHR00 also offer a prescription for determining $\lambda_i$, the mean free path of ionizing photons.  In their model, it equals the mean distance between clumps with $\Delta>\Delta_i$ along a random line of sight, which is approximately
\bq
\lambda_i = \lambda_0 [1 - F_V(\Delta_i)]^{-2/3}.
\label{eq:mfp-mhr}
\eq
Here $F_V(\Delta_i)$ is the fraction of volume with $\Delta < \Delta_i$ and $\lambda_0$ is a (redshift-dependent) normalization factor.  Formally, this expression is valid only if the number density and shape (though not total cross section) of absorbers is independent of $\Delta_i$.  This is obviously not true in detail for the cosmic web.  However, MHR00 found that it provided a good fit to numerical simulations at $z=2$--$4$ if we set $\lambda_0 H(z) = 60 \kms$ (in physical units).  

\subsection{The Recombination Limit} \label{recomb-limit}

For simplicity, we will assume that gas with $\Delta<\Delta_i$ is highly ionized while all gas with $\Delta > \Delta_i$ remains neutral, so that $\lambda_i$ is an accurate approximation to the mean free path.  This provides a reasonable description of shielding in dense regions, if those regions can be considered to be isolated clumps in which the density increases inwards (see the Appendix to FO05 for more detail).  In that case, the radiation field will ionize the outskirts of the cloud until $\tau \approx 1$.  Because of the density gradient (increasing inwards), this skin corresponds to our threshold $\Delta_i$.  In detail, equation (\ref{eq:mfp-mhr}) ignores two significant (but opposing) effects.  First, it probably overestimates $\lambda_i$ by up to a factor $\sim 2$ because of accumulated photoelectric absorption by low column density systems (FO05).    But second, it ignores higher energy photons, which are able to travel farther.  We will assume that these two effects roughly cancel.

To obtain $R_{\rm max}$, the maximum bubble size in the presence of recombinations, we simply find the point at which the  mean free path $\lambda_i(\Delta_i)$ equals the distance from the quasar.\footnote{This procedure makes the accumulated opacity of low-column density systems (with $\Delta < \Delta_i$) less important, because $\Gamma \propto 1/R^2$ along each line of sight.  As a result, most of the absorption occurs near $R_{\rm max}$, so the factor of two uncertainty in the mean free path from weak systems decreases $R_{\rm max}$ by a significantly smaller amount.}  The filled triangles in Figure~\ref{fig:DeltaR} show $R_{\rm max}$ for several quasar luminosities.  It ranges from $\sim 5$--$100 \Mpc$ over this luminosity range.  For the faintest quasars, $\Delta_i \la 5$; in these cases, the assumption of a sharp transition in the neutral fraction at $\Delta_i$ is probably not a good one, because the gas is so close to the mean density.  However, at $L \ga 10^{11} \Lsun$, $\Delta_i \ga 8$ and $F_V(\Delta_i) < 0.03$, so the two-phase approximation is probably reasonable.

There is one important caveat about this method:  the MHR00 procedure computes the mean free path at an arbitrary point in the IGM, whereas we are interested in the mean free path as seen from the quasar, which most likely sits in an overdense regions (see also \citealt{yu05, alvarez07, lidz07}).  Thus we probably overestimate the mean free path.  However, even at $z \sim 6$ (where these massive halos are much more rare) the environments typically approach the mean density within $\la 20 \Mpc$ of the quasar, so this should not be a huge effect.

\begin{figure}
\plotone{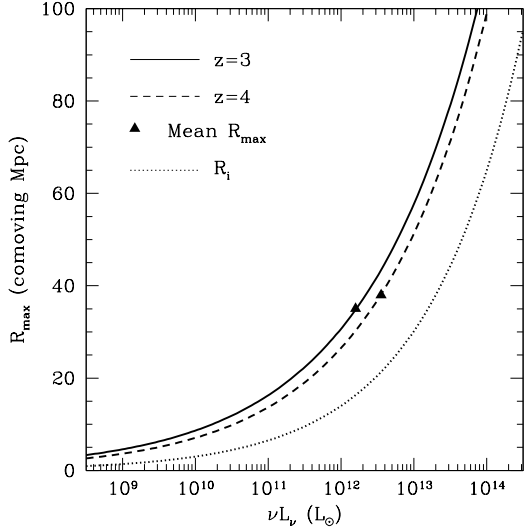}
\caption{Radius at which $R=\lambda(\Delta_i)$ (i.e., the point at which the expected IGM mean free path equals the distance from the quasar), as a function of quasar luminosity.  The solid and dashed curves are for $z=3$ and $4$, respectively.  The filled triangles mark the luminosity-weighted mean $R_{\rm max}$ for each case.  The dotted curve shows the maximum radius for an ionized bubble around an isolated source.}
\label{fig:rmax}
\end{figure}

Figure~\ref{fig:rmax} shows $R_{\rm max}$ as a function of luminosity for quasars at $z=3$ and $4$ (solid and dashed curves, respectively).  The filled triangles indicate the luminosity-weighted mean $R_{\rm max}$ across the entire quasar population.  The dotted curve shows $R_i$ from equation~(\ref{eq:Vi}), the maximum size of the ionized bubble surrounding an isolated source (assuming $t_7=1$, $\bar{x}_{\rm HeII}^u=1$, and $f_{\rm abs}=1$).  We find $\VEV{R_{\rm max}} \approx 35$--$37 \Mpc$ over this redshift range; the increasing clumpiness and increasing mean luminosity roughly cancel each other out.  Crucially, $R_{\rm max} > R_i$ for all luminosities, so that isolated quasars will not be seriously affected by recombinations.  However, the limit is also not extremely large, and it will clearly come into play for quasars that appear inside of large pre-ionized bubbles, preventing them from contributing as much to reionization as they otherwise would.  Comparing to Figure~\ref{fig:rchar}, recombinations will start to limit the bubble growth when $\bxion \sim 0.6$--0.8.

\subsection{The Bubble Distribution with Recombinations}
\label{bubrec}

Finally, we are ready to compute the bubble size distribution when the recombination limit is included.  In that case, we approximate the excursion set barrier as the usual version for $R<R_{\rm max}$.  There is no barrier at all for $R>R_{\rm max}$ (so no such bubbles can form), and at $R=R_{\rm max}$ the barrier is simply a vertical line from the edge of the usual version to positive infinity.  Although this is obviously approximate, recombinations do become important very quickly for a given quasar luminosity, so it is an excellent model (but see below).  With it, we can solve for the bubble distribution analytically (FO05).  Because there is no barrier at $R>R_{\rm max}$, the probability distribution at that size scale is a simple gaussian:
\bq
p(\delta | R_{\rm max}) \, \deriv \delta = \frac{1}{\sqrt{2 \pi} \sigma_{\rm max}} \ \exp \left( \frac{-\delta^2}{2 \sigma_{\rm max}^2} \right) \, \deriv \delta,
\label{eq:pdel}
\eq  
where $\sigma_{\rm max} \equiv \sigma(R_{\rm max})$.  The modified barrier is a vertical line beginning at $B(R_{\rm max})$, so any trajectory that has $\delta(R_{\rm max})>B(R_{\rm max})$ will be incorporated into a bubble at precisely the limiting radius.  Their number density is
\bq
N_{\rm rec}(m_{\rm max}) =  \frac{\bar{\rho}}{2 m_{\rm max}} \ {\rm erfc} \left[ \frac{B(R_{\rm max})}{\sqrt{2} \sigma_{\rm max}} \right]
\label{eq:fcl}
\eq
(note that this is a true number density with units inverse comoving volume, not a density per unit bubble radius).

Trajectories with $\delta(R_{\rm max})<B(R_{\rm max})$ continue their random walks until they cross the photon-counting barrier on smaller scales.  The result must be independent of the trajectory at $R>R_{\rm max}$; we only care about its value where the barrier begins.  The mass function is [excluding the true recombination-limited bubbles of eq. (\ref{eq:fcl})]:
\bq
n_{\rm rec}(m,z) =  \int_{-\infty}^{B(R_{\rm max})} \deriv \delta \ p(\delta | R_{\rm max}) \ n_b(m,z|\delta,R_{\rm max},z),
\label{eq:nmrec}
\eq
where $n_b(m,z|\delta,R_{\rm max},z)$ is the conditional mass function for a trajectory that begins its random walk at the point $(\sigma_{\rm max}^2,\delta)$.  In other words, the net mass function is the weighted average of the conditional mass functions evaluated over all densities smoothed on the scale $R_{\rm max}$.

\begin{figure}
\plotone{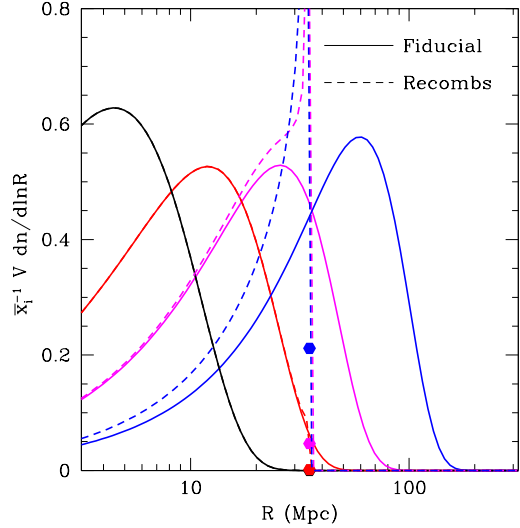}
\caption{Bubble size distributions at $z=3$ with and without recombinations (dashed and solid curves, respectively).  Both sets assume a constant ionizing efficiency and $\mmin=10 m_i$.  In each case, the curves have $\bxion=0.2,\,0.4,\,0.6$, and $0.8$, from left to right.  The filled hexagons show the fraction of space in bubbles with $R=R_{\rm max}$ (see text); the $\bxion=0.2$ and $0.4$ hexagons overlap near zero.}
\label{fig:nbub_rec}
\end{figure}

Although the integral in equation (\ref{eq:nmrec}) can be solved analytically (FO05), the result is complicated and far from illuminating.  We therefore simply show some example size distributions in Figure~\ref{fig:nbub_rec}.  The dashed and solid curves are $n_b$ with and without recombinations, respectively.  The four sets take $\bxion=0.2,\,0.4,\,0.6$, and $0.8$, from left to right.  The behavior is fairly similar to the analogous case of hydrogen reionization:  recombinations are relatively unimportant for small $\bxion$, but they impose a sharp maximum on the bubble size distribution when $\bxion \ga 0.5$.  As expected, the limit sets in somewhat earlier for helium, because of its enhanced recombination rate and the increased clumpiness at later times.  

Unfortunately, a comparison of Figures~\ref{fig:nbub} and \ref{fig:nbub_rec} shows that recombinations limit bubble sizes to only a few times the characteristic scale from stochastic reionization.  Thus they will seriously limit the range of $\bxion$ for which density-driven reionization could be observable.

There is one significant problem with our use of a single $R_{\rm max}$ value across all bubbles.  During hydrogen reionization, this is an excellent approximation; each bubble has hundreds of sources by this stage, so fluctuations in the emissivity (and hence recombination limit) are small \citep{furl05-charsize}.  However, we saw in \S \ref{recomb-limit} that with helium reionization we generally have only a single active ionizing source per bubble, and $R_{\rm max}$ is sensitive to its luminosity (changing by roughly a factor of two for each decade in luminosity).  Thus some bubbles, which contain multiple luminous quasars or a single very luminous quasar, will be able to grow well beyond the $R_{\rm max}$ shown in Figure~\ref{fig:nbub_rec}.  This is simply another manifestation of our expectation that while the morphology of hydrogen reionization should be driven primarily by the clustering of ionizing sources, stochastic fluctuations are much more important in helium reionization (and can continue to play a role late in the process).  Although these bright sources cause only a relatively small fraction of ionizations, they will certainly make the recombination limit more of a smooth cutoff than a hard barrier.

In principle, this formalism allows us to compute the effective clumping factor in quasar bubbles, and hence the evolving recombination rate throughout reionization (see FO05 for a similar calculation during hydrogen reionization).  One could then compare to the empirical ionization histories in \S \ref{empirical} and constrain the importance of recombinations.  In practice, this is rather difficult, because most of the recombinations actually occur during fossil phases (see \S \ref{fossils} below), because we have neglected clustering of absorbers around the quasar, and because we lack a well-motivated model for the quasar hosts.  However, this model \emph{can} be tested if one can obtain an independent measure of the \ion{He}{3} ionizing background.  Because we know the emissivity from the luminosity function, this would then constrain the mean free path -- and hence implicitly the recombination rate and our saturation radius.  

\subsection{A Consistency Check:  The Post-Overlap IGM} \label{post-overlap}

As mentioned above, this recombination limit controls the transition to the post-overlap IGM, so it is useful to consider observational constraints on the helium-ionizing photon mean free path in that regime.  While our direct knowledge of the helium \lya forest is limited, we can make a number of predictions about its post-reionization properties from the hydrogen \lya forest.  In the optically thin limit, the helium and hydrogen \citet{gunn65} optical depths are \citep{miralda93}
\bq
{\tau_{\rm GP,HeII} \over \tau_{\rm GP,HI}} = {\eta_{\rm thin} \over 4},
\label{eq:taucomp}
\eq
where $\eta_{\rm thin} = N_{\rm HeII}/N_{\rm HI}$ and $N_x$ is the column density of species $x$.  Observations show a wide scatter in the values of $\eta_{\rm thin}$ in individual systems, but the median is $\sim 45$--$80$ \citep{shull04,zheng04}.  The \citet{haardt96} model for the ionizing background predicts that $\VEV{\eta_{\rm thin}} \sim 40$ in a fully-ionized universe at moderate redshifts, while \citet{fardal98} predict $\VEV{\eta_{\rm thin}} \sim 80$.  

In the optically thick regime, a more careful calculation including radiative transfer is required.  If we define $N_{\rm LLS}$ to be the column density of neutral hydrogen for which $\tau=1$ at the Lyman-edge of hydrogen, and $N_{\rm HeLLS}$ to be the column density of neutral hydrogen for which $\tau=1$ at the Lyman-edge of helium, the optically thin form of equation~(\ref{eq:taucomp}) would suggest $N_{\rm HeLLS} \approx \kappa 4 N_{\rm LLS}/\eta_{\rm thin}$.  Radiative transfer calculations of slabs illuminated from both sides then suggest $\kappa \sim 2$ \citep{haardt96}.  Thus, absorbers that are opaque to helium-ionizing radiation have column densities $\sim 0.1$--$0.2 N_{\rm LLS}$.  The distribution of \hone absorbers goes like $f \propto N_{\rm HI}^{-1.5}$ \citep{petitjean93}.  Therefore, if these opaque systems were entirely responsible for the absorption, we would expect the mean free path of helium-ionizing photons to be $\lambda_{\rm He} \sim 0.03$--$0.1 \lambda_{\rm H}$.  

This is useful because the properties of hydrogen-ionizing photons are well-studied.  \citet{storrie94} found that the abundance of Lyman-limit systems is $dN_{\rm LLS}/dz \approx 3.3 [(1+z)/5]^{3/2}$.   The mean free path for a photon at the hydrogen Lyman-edge is therefore \citep{miralda03}
\bq
\lambda_{\rm H} \approx 110 \left( {5 \over 1+z} \right)^3 \Mpc
\eq
in comoving units, where we have again assumed $f \propto N_{\rm HI}^{-1.5}$.  Thus $\lambda_{\rm He} \sim 6.6,\,12,$ and $30 \Mpc$ at $z=4,\,3,$ and 2, with about a factor of two uncertainty from the spread in $\eta_{\rm thin}$.  These are somewhat smaller than our $R_{\rm max}$ values, but that is not surprising:  even after overlap, most of the universe does not lie near an active, bright quasar, and the mean free path will fluctuate depending on whether an active source is nearby.  Other estimates of the attenuation length around active quasars after overlap are comparable to ours (e.g., \citealt{bolton06} take $R_{\rm max}=30 \Mpc$).

\subsection{Helium Bubbles in the Clumpy IGM} \label{bub-clump}

As described in \S \ref{blur}, quasars have rather hard spectra, which blurs the edges of ionized bubbles (and allows some photons to escape to infinity).  Previously we estimated the thickness of these bubbles assuming a uniform IGM; of course, a more accurate treatment includes the discrete absorbers discussed above.

In detail, if the number of absorbers with column density $N_{\rm HeII}$ per unit redshift is $f(N_{\rm HeII},z)$, the effective optical depth experienced by a photon as it travels through a redshift interval $(z_1,z_2)$ is \citep{zuo93}
\bq
\tau_{\rm eff} = \int_{z_1}^{z_2} \deriv z \int \deriv N_{\rm HeII} f(N_{\rm HeII},z) [ 1 - e^{-N_{\rm HeII} \sigma_{\rm HeII}(\nu)} ].
\label{eq:teff-fn}
\eq
If we then assume that $f(N_{\rm HeII}) \propto N_{\rm HeII}^{-3/2}$ (so that the helium clumps trace the \ion{H}{1} \lya forest; \citealt{rauch98}), we find $\lambda \propto \nu^{-3/2}$, a much gentler increase with frequency than for a uniform IGM.  In this case, half the ionizing photons have mean free paths smaller than $\approx 1.9$ times the value at the ionization edge (as opposed to a factor $\approx 3.6$  for a uniform IGM).  Of course, the power law appropriate for the \lya forest may not be applicable before helium reionization is complete, but it shows that clumping in the  neutral helium phase will help sharpen the bubbles and hence make our model more accurate.

\subsection{Recombinations Between Quasar Episodes} \label{fossils}

To this point, we have only discussed recombinations while a quasar is active.  Most likely, this period fills only a small fraction of the age of the universe, and a substantial amount of helium can actually recombine during the long dormant phases.  We explore this aspect more fully in \citet{furl08-fossil}; here we confine ourselves to a few words about their effect on the morphology.

As shown in equation~(\ref{eq:rectime}), the recombination time of mean density gas is smaller than the Hubble time throughout helium reionization.  The recombination rate will be enhanced by clumping:  the MHR00 model predicts that $C \approx 3$--$4$ during this era, if all the non-virialized gas is fully ionized.  However, as described in \citet{furl08-fossil}, the clumping enhancement is transient during a recombination phase:  once the densest gas has fully recombined, it can no longer aid in future recombinations.  Thus the \emph{effective} clumping for recombinations quickly falls near unity.  

Moreover, because helium reionization occurs so quickly (with $\sim 75\%$ of the ionizations at $z<4$), there is relatively little time available between quasar generations for recombinations to occur -- typically $\la 10\%$ of the Hubble time, except for very high redshift bubbles \citep{furl08-fossil}.  Over this time span, less than $\sim 40\%$ of the gas is able to recombine, even with clumping included, and that fills only $\sim 10\%$ of the volume.  In terms of an isolated bubble, this means that high-density filaments and sheets will mostly recombine before the next quasar illuminates the region, but most of the \emph{volume} will remain ionized, and the morphology will be largely unaffected -- although the next generation of quasars will have to reionize these dense pockets.

On the other hand, ionized regions generated by rare, extremely high-redshift quasars will recombine significantly before being illuminated again; as a consequence, many of their photons are essentially wasted during this early phase.

\section{Discussion} \label{disc}

We have examined several aspects of helium reionization using simple analytic models.  We first showed that the \citet{hopkins07} luminosity function, together with standard template quasar spectra, produces $\sim 2.5$ ionizing photons per helium atom at $z=3$, nicely consistent with observational hints that the IGM truly becomes fully ionized at that time.  Indeed, even with an average clumping factor $\bar{C}=3$, the known quasar population can ionize helium by that time.  Interestingly, the quasar emissivity increases rapidly at $z \ga 3$, with $\sim 75\%$ of the ionizing photons produced after $z=4$.  This is somewhat faster than naive models based on the collapse fraction predict, implying that supermassive black holes in quite massive halos dominate the photon budget.

In our calculations, we have used a quasar luminosity function with a shallow faint-end slope at $z \ga 3$ \citep{hunt04, cristiani04, hopkins07}.  This implies that most ionizing photons are produced by bright quasars ($L \ga L_\star$), which are of course relatively rare.  As a result, the initial phases of reionization are dominated by rare, large bubbles (with characteristic size $R_{\rm char} \sim 15 \Mpc$)  whose overlap is random.  We described this regime with a ``stochastic" model where sources are distributed according to Poisson statistics.  In this model, the bubble size distribution follows a power law (roughly), so there is no true characteristic scale -- although the power law is so steep that, in practice, nearly all bubbles contain one or at most a few sources anyway.  Thus, in the early stages of reionization, we expect bubbles to have a characteristic size corresponding to a single or at most a few sources.  However, more recent data shows a somewhat steeper faint-end slope \citep{fontanot07, bongiorno07, siana08}.  This leads to a broader distribution of luminosities, making the stochastic phase somewhat less important.  

Large-scale clustering of the ionizing sources will eventually cause the bubbles to grow beyond this point.  We use the model of FZH04 to describe this phase, in which large-scale overdensities are ionized first.  This imprints a well-defined characteristic scale, which first exceeds the size of the stochastic bubbles when $\bxion \sim 0.5$ and would exceed $100 \Mpc$ by the end of reionization if allowed to grow unchecked.  The scale is nearly independent of redshift at a fixed $\bxion$, but it does depend on the clustering of the quasar hosts, with more massive host galaxies driving larger bubbles.
In our calculations, we used a simple halo-based model to represent quasar hosts.  This is reasonable if the observed relations between black hole mass and halo properties \citep{magorrian98, gebhardt00, ferrarese00} hold at high redshift as well.  But many detailed models of the quasar luminosity function rely on mergers to trigger quasar phases.  This would change the distribution of quasar luminosities over halos and so change some of the predictions of our model (see also \citealt{cohn07}).  We intend to examine these effects with the semi-numeric approach of \citet{mesinger07}.

However, even during this phase nearly all quasars exist in isolation:  provided that their duty cycle $t_{\rm QSO} H(z) \ll 1$, very few bubbles will contain more than one \emph{active} source.  Thus the ionizing background around each source is easy to compute (modulo the nearly uniform background of high-energy photons with large mean free paths).  Together with a model for the IGM, this allows us to estimate the mean free path of helium-ionizing photons as they propagate through the ionized bubbles and specifically to compute how far they can travel before being consumed by dense, optically thick absorbers.  We find that, at the ionization edge, $R_{\rm max} \sim 35$--$40 \Mpc$ when averaged over the entire quasar population.  Once bubbles grow to this size, only a fraction of the photons would ionize new helium atoms.  As described in FO05, the effective clumping factor of the IGM will increase as more and more photons get consumed by these dense systems, so reionization will slow down.  Eventually, the ionized gas distribution will map onto the more uniform background characteristic of the post-reionization era (although even then large fluctuations will persist; \citealt{shull04}).

Our model does not include recombinations within bubbles that contain zero active quasars over a period of time, except as an overall scaling factor to the the ionizing efficiency (through $\bar{N}_{\rm rec}$).  The helium recombination time is short enough, and the IGM clumpiness large enough, that a significant fraction of helium atoms (mostly in high density regions) can recombine, although $\ga 90\%$ of the volume will remain highly ionized.  We examine these ``fossil" bubbles more closely in \citet{furl08-fossil}.

Compared to hydrogen reionization, the helium era passes through the same qualitative phases: stochastic, isolated bubbles followed by a rapid increase in the bubble size when clustering becomes significant and finally a ``saturation" when the mean free path of an ionizing photon becomes smaller than the bubble size.  However, quasars are much more rare than star-forming galaxies at $z \ga 6$ (at least in most models of the high-redshift universe), so the early stochastic phase is much more important; for hydrogen, it ends by $\bxion \sim 0.1$.  There is thus a much shorter window in which density-driven clustering dominates the morphology, when $R_{\rm stoch} < R < R_{\rm max}$.  

Because there is typically no more than one quasar per bubble, the scale $R_{\rm max}$ is much easier to estimate than for hydrogen.  FO05 estimated it during hydrogen reionization by balancing the total recombination rate within the bubble against the emissivity of the ionizing sources.  This depends on the recombination history and uncertain IGM density distribution, and it assumes that, within each bubble, the ionization always begins at low densities.  As a result, $R_{\rm max}$ is quite sensitive to many of the input parameters (varying by about a factor of two between case-A and case-B recombination, for example).   

For helium reionization, $R_{\rm max}$ can be much more robustly determined because it depends only on ionization balance around a source population (whose luminosity function is quite well-known, at least at $z \la 4$) in an IGM whose density structure is extremely well-determined by the \ion{H}{1} \lya forest (although the spread in luminosities also makes it more difficult to incorporate directly into the simple analytic model).  The helium reionization era will thus offer a much more stringent test of the transition to the recombination-dominated phase than the hydrogen era.  

There is, however, a possibility that hydrogen and helium reionization are more similar than it may appear.  We showed that observations at least allow the possibility that $z \approx 3$ galaxies produce nearly as many helium-ionizing photons as quasars.  Their importance for reionization depends entirely on their escape fraction, which is difficult to estimate.  If they do contribute, we would expect a slower evolution of $\bxion$ with redshift, a softer ionizing background, many more small ionized bubbles, and a more uniform, smaller mean free path (and hence $R_{\rm max}$) for the helium-ionizing photons.

We have focused exclusively on analytic models here, but of course a full picture of helium reionization requires numerical simulations that capture both the stochastic and large-scale clustering elements of our model, as well as the complex geometry of the IGM and source distribution.  The models described here can be extended to ``hybrid" semi-numeric schemes \citep{zahn07-comp, mesinger07} and also full radiative transfer simulations (e.g., \citealt{sokasian03}).  However, some elements -- principally recombinations -- will require either small-scale hydrodynamic simulations or analytic prescriptions similar to the ones used here to model accurately.  Our analytic models thus provide a basic underpinning for understanding and improving more complex numerical realizations of reionization.  

Finally, we must acknowledge that observing the morphology of helium bubbles directly will likely be rather difficult.  During hydrogen reionization, the two most promising techniques are to use the 21 cm transition of neutral hydrogen to map the bubble distribution (e.g., \citealt{furl06-review}) and to infer the distribution of neutral gas from \lya line-selected galaxies.  Neither is available here, because $^4$He lacks a hyperfine transition and galaxies are already so faint in the far-UV.  Instead we must turn to more subtle approaches.  One possibility is through \ion{He}{2} \lya forest absorption spectra of distant quasars:  large bubbles may manifest themselves as transmission spikes separated by long ``dark gaps" of saturated absorption.  Another is to search for large-scale coherent fluctuations of some proxy for the ionized helium abundance, such as in the thermal properties of the IGM \citep{furl07-igmtemp}, absorption in the \ion{H}{1} \lya forest, or the abundance ratios of metal ions.  We intend to explore all of these issues in future work.

\acknowledgments

We thank Z. Haiman, L. Hernquist, A. Lidz, M. McQuinn, J. Schaye, B. Siana, and M. Zaldarriaga for helpful comments.  Figure~\ref{fig:qlf} was generated with the quasar luminosity function script of \citet{hopkins07}.  SRF acknowledges support from NSF grant AST-0607470. SPO acknowledges NSF grant AST-0407084 and NASA grant NNG06H95G for support. 

\bibliographystyle{apj}
\bibliography{Ref_composite}

\end{document}